\documentclass[letterpaper, 12pt, onecolumn]{IEEEtran}
\usepackage[unicode]{hyperref}
\usepackage{amsmath} 
\usepackage{amssymb} 
\usepackage{amscd}
\usepackage{courier}
\usepackage{lastpage}
\usepackage{graphics}

\newtheorem{lemma}{Lemma}
\newtheorem{theorem}{Theorem}
\newtheorem{property}{Property}
\newtheorem{definition}{Definition}

\def \rve {\mathop {\mathbb{E} }}

\def \argmax {\mathop {\rm arg\ max }}

\def \Pr {{\mathbb{P}}}

\title{Improved Source Coding Exponents \\ via Witsenhausen's Rate}

\author{\IEEEauthorblockN{Benjamin G. Kelly and Aaron B. Wagner}
\IEEEauthorblockA{\\School of Electrical and Computer Engineering\\
Cornell University\\
Ithaca, NY 14853\\
bgk6@cornell.edu, wagner@ece.cornell.edu}}

\begin{document}

\maketitle 

\begin{abstract}
	We provide a novel upper-bound on Witsenhausen's rate, the rate required in the zero-error analogue of the Slepian-Wolf problem; our bound is given in terms of a new information-theoretic functional defined on a certain graph.  We then use the functional to give a single letter lower-bound on the error exponent for the Slepian-Wolf problem under the vanishing error probability criterion, where the decoder has full (i.e. unencoded) side information.  Our exponent stems from our new encoding scheme which makes use of source distribution only through the positions of the zeros in the `channel' matrix connecting the source with the side information, and in this sense is `semi-universal'.  We demonstrate that our error exponent can beat the `expurgated' source-coding exponent of Csisz\'{a}r and K\"{o}rner, achievability of which requires the use of a non-universal maximum-likelihood decoder.  An extension of our scheme to the lossy case (i.e. Wyner-Ziv) is given.  For the case when the side information is a deterministic function of the source, the exponent of our improved scheme agrees with the sphere-packing bound exactly (thus determining the reliability function).  An application of our functional to zero-error channel capacity is also given.
\end{abstract}

\section{Introduction}

\begin{figure}
\centering
\scalebox{0.5}{\includegraphics*{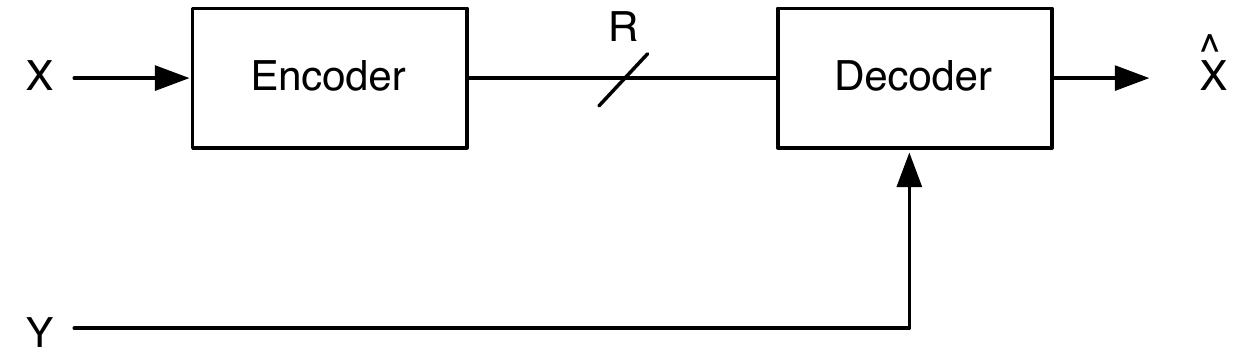}}
\caption{Source coding with full side information}
\label{fig:scsi}
\end{figure}

Under consideration is the communication problem depicted in Figure \ref{fig:scsi}; nature produces a sequence $(X_i,Y_i)$ governed by the i.i.d. distribution $P_{XY}$ on alphabet $\mathcal{X} \times \mathcal{Y}$.   An encoder, observing the sequence $X^n$, must send a message to a decoder, observing the sequence $Y^n$ (the side information), so that the decoder can use the message and its observation to generate $\hat X^n$, an estimation of $X^n$ to some desired fidelity. 

For lossless reproduction, using the criterion that $P_{XY}^n(X^n \ne \hat X^n) \to 0$ as the blocklength $n \to \infty$, Slepian and Wolf \cite{Slepian:1973p333} determined that all rates in excess of $H(X|Y)$ are achievable.  Bounds on the rate of decay of the error probability for this problem, the so-called \emph{error exponent},  were determined by Csisz\'{a}r and K\"{o}rner \cite{Csiszar:1981p220} whose results include a universally attainable random coding exponent and a non-universal `expurgated' exponent.  Previously Gallager \cite{gallagersideinfo} derived a non-universal exponent that was later shown to be universally attainable by Csisz\'{a}r, K\"{o}rner and Marton \cite{Csiszar:1980p344}.  For the Slepian-Wolf problem in its full generality (i.e. allowing for coded side information) the best known exponents are those of Csisz\'{a}r \cite{Csiszar:1982p334} and Oohama and Han \cite{Oohama:1994p222}.  In the regime where the rate of the second encoder is large, our new exponent also improves upon these results, but we do not consider the general case here.

In the case of lossy reproduction, with the loss measured by some single letter distortion function $d$, the scenario is known as the Wyner-Ziv problem \cite{wynerziv}, after Wyner and Ziv who showed that if the allowable expected distortion is $\Delta$, then the required rate is given by
$$
R_{WZ}(P_{XY},\Delta)=\inf I(X;U) - I(Y;U),
$$
where the infimum is over all auxiliary random variables $U$ such that (1) $U$,
$X$, and $Y$ form a Markov chain in this order and (2) there exists a function
$\phi$ such that
$$
\rve[d(X,\phi(Y,U))] \le \Delta.
$$
The best available exponents for the Wyner-Ziv problem were determined by the present authors in \cite{kellywagnerscexponents}.  Henceforth we refer to both lossless and lossy problems as \emph{full side information problems}.

We describe new encoding schemes for both full side information problems which rely on ideas from graph theory.  Our analysis shows that the chromatic number of a particular graph can be used to characterize the number of sequences that can be communicated without error.  We are able to give a single letter upper bound on this chromatic number via a new functional on a graph $G$. We call our schemes \emph{semi-universal} because the scheme depends on the source distribution only through the position of the zeroes in the channel matrix.  By comparing our new exponent directly with the previous results one sees that our scheme is capable of sending a larger number of sequences without error, i.e. we can expurgate more types which leads to better exponents.  

Although our scheme applies to the vanishing error probability case, it is derived from the study of a related zero-error problem.  The zero-error formulation of source coding with full side information was studied by Witsenhausen \cite{Witsenhausen:1976p290}, who showed that for fixed blocklength, $n$, the fewest number of messages required so that the decoder can reproduce the source with no error, i.e. $P_{XY}^n(X^n = \hat X^n)=1$, is $\gamma(G_X^n)$, the chromatic number of the $n$-fold strong product of the characteristic graph of the source.

The required rate, sometimes referred to as Witsenhausen's rate in the literature, is therefore
\begin{equation}
\label{wrrate}
R(G) = \lim_{n \to \infty} \frac{1}{n} \log \gamma(G^n).
\end{equation}
(We note that limit in \eqref{wrrate} exists by sub-additivity and appealing to Fekete's lemma.)  Unfortunately, the problem of determining $R(G)$ `seems, in general, far beyond the reach of existing techniques' \cite{alon:powers}; see also the comment at the end of section \ref{sect:wr} here.  However, since $\gamma(G^n) \le \gamma(G)^n$, it is clear that $R(G) \le \log \gamma(G)$.  We provide a new bound on $R(G)$ by bounding the chromatic number of $G^n$ restricted to typeclasses.  Our techniques combine graph- and information-theoretic techniques, see K\"{o}rner and Orlitsky \cite{Korner:1998p65} for a comprehensive overview of the applications of graph theory in zero-error information theory. 

The rest of the paper is organized as follows.  Section \ref{sect:def} gives definitions. Section \ref{sect:kappaprops} gives some useful properties of $\kappa$.  In Section \ref{sect:wr} we motivate $\kappa$ and give our first result, a single letter bound on Witsenhausen's rate. In Section \ref{sect:exp}, we give our second result, improved error exponents for the problem of lossless source coding with full side-information; examples and comparisons to previous known exponents are also given.  In Section \ref{sect:coveringbinning} we use the ideas from Section \ref{sect:exp} to give our third and fourth results, an improved error exponent for the lossy problem and determination of the reliability function for the case when the side information is a deterministic function of the source.  In Section \ref{sect:channelcoding} we briefly give an application of $\kappa$ to channel coding.

\section{Definitions}
\label{sect:def}

Script letters, e.g. $\mathcal{X}, \mathcal{Y}$, denote alphabets.  The set of all probability distributions over an alphabet $\mathcal{X}$ will be denoted by $\mathcal{P}(\mathcal{X})$.  Small bold-faced letters, e.g. $\mathbf{x} \in \mathcal{X}^n, \mathbf{y} \in \mathcal{Y}^n$ denote vectors, usually the alphabet and length are clear from the context.  For information-theoretic quantities, we use the notations of \cite{ckit}.  $H(\mathbf{x} | \mathbf{y})$ denotes conditional empirical entropy, i.e. the conditional entropy computed using the empirical distribution $P_{\mathbf{x},\mathbf{y}}$.  We use $[x]^+$ to denote $\max(0,x)$.  Unless specified, exponents and logarithms are taken in base 2.

A graph $G=(V,E)$ is a pair of sets, where $V$ is the set of vertices and $E \subset V \times V$ is the set of edges.  Two vertices $x,y \in V$ are connected iff $(x,y) \in E$. We will restrict ourselves to simple graphs, i.e. undirected graphs without self-loops.  The \emph{degree of a vertex} $v$, $\Delta(v)$, is the number of other vertices to which $v$ is connected.  The \emph{degree of a graph} $G$, denoted $\Delta(G)$ is defined as $ \max_{v \in V} \Delta(v)$.  A coloring of a graph is an assignment of colors to vertices so that no pair of adjacent vertices share the same color.  The \emph{chromatic number} of $G$, $\gamma(G)$, is defined to be the fewest number of colors needed to color $G$.  For $U \subset V$, $G(U)$ is the \emph{(vertex-) induced subgraph}, i.e. the graph with vertex set $U$ and edge set $E \cap (U \times U)$.  For two matrices, $V, W$ we use $V \ll W$ to mean that $W(b|a)=0$ implies $V(b|a)=0$.

Let $G=(V,E), H=(V',E')$ be two graphs. The \emph{strong product} (or \emph{and product}) $G \wedge H$ is a graph whose vertex set is $V \times V'$ and in which two vertices $(v,v'), (u,u')$ are connected iff
\begin{enumerate}
	\item $v=u$ and $(v',u') \in E'$ or  
	\item $v'=u'$ and $(v,u) \in E$ or
	\item $(v,u) \in E$ and $(v',u') \in E'$. 
\end{enumerate}
We will be interested in $G^n=G \wedge G \wedge \ldots \wedge G$ ($n$-factors), the $n$-fold strong product of $G$.  One may think of the vertices of $G^n$ as length $n$ vectors $(v_1, \ldots, v_n)$ with two vertices are connected in $G^n$ if all of the components of the vectors are the same or connected in $G$.  The \emph{characteristic graph}, $G_X$, of a source $P_{XY}$ is the graph whose vertex set is $\mathcal{X}$ and two vertices $x, x'$ are connected if there is a $y \in \mathcal{Y}$ such that $P(y|x')P(y|x) >0$.  For a given $\mathbf{y}$, the set $Z(\mathbf{y}) = \{\mathbf{x} : P(\mathbf{x} | \mathbf{y}) > 0 \}$ is the set of `confusable' sequences, i.e. the set of $\mathbf{x}$s than can occur with a given $\mathbf{y}$.   For a graph $G$ and distribution $Q$ on the vertices of $G$, we define the following functional.
\begin{definition}
\begin{align}
 \kappa(G, Q) = \mathop{\max_{V: V \ll G}}_{Q V = Q} H(V | Q).
\end{align}
Note when we write the graph $G$ where a matrix is expected, we abuse notation and refer to the matrix $G=A+I$ where $A$ is the adjacency matrix of graph $G$ and $I$ is the identity matrix.

Equivalently one may think of $\kappa$ as follows
\[
\kappa(G, Q) = \mathop{\max_{X, \tilde X:}}_{Q_X = Q_{\tilde X}=Q} H(\tilde X | X).
\]
where $X$ and $\tilde X$ have common alphabet and $P(\tilde x | x) >0$ iff $\tilde x, x \in E(G)$.

\end{definition}

\section{Properties of \texorpdfstring{$\kappa$}{\textkappa}}
\label{sect:kappaprops}
In this section we give some properties of $\kappa$ which will be used elsewhere in the paper.  Throughout this section $G$ is a graph, $Q$ is a distribution on the vertices of $G$ and $X$ is a random variable with distribution $Q$.
\begin{property}
\label{prop:kapent}
$\kappa(G, Q) \leq H(Q)=H(X)$, where equality holds if $G$ is fully connected.
\end{property}
\begin{IEEEproof}
Note that any valid choice of channel in the optimization defining $\kappa(G,Q)$ satisfies $Q V = Q$, thus $H(V|Q) \le H(Q)$, giving the first claim.

If $G$ is fully connected then the constraint $V \ll G$ imposes no restriction on the choice of $V$.  The problem is then to choose a $V$ that produces the given output distribution $Q$.  Setting the rows of $V$ equal to $Q$ gives $\kappa(G,Q)=H(Q)$.  
\end{IEEEproof}
\begin{property}
\label{prop:hxcy}
If $G$ is the disjoint union of fully connected subgraphs then
\begin{equation}
	\kappa(G,Q) = H(X|Y).
\end{equation}
where 
\begin{enumerate}
	\item $Y$ is a random variable with alphabet size $|\mathcal{Y}|$ equal to the number of disjoint subgraphs in $G$ so that to each subgraph we associate a unique element $y \in \mathcal{Y}$; and
	\item for the subgraph associated with $y$, the event $\{X=a, Y=y\}$ has probability $Q(a)$ if $a$ is in the subgraph and probability zero otherwise.
\end{enumerate}
\end{property}
\begin{IEEEproof}
	Without loss of generality we may assume the adjacency matrix of $G$ plus the identity matrix is block diagonal, where each block corresponds to a fully connected subgraph (i.e. is all 1s).  By independence it suffices to solve the maximization problem for one of these blocks, say the one associated with element $y$.  
	
  Suppose that the subgraph has vertices $a_1, a_2, \ldots, a_n $ and define the (semi) probability measure $Q_y = [Q(a_1)~Q(a_2)~\ldots~Q(a_n)]$.  Then the problem is
\begin{equation}
\label{eqn1:hxcystep1}
	\max_{V: Q_y V = Q_y} \sum_{a} Q_y(a) \sum_{b}  -V(b|a) \log V(b|a).
\end{equation}
Let $\tilde Q_y = \frac{Q_y}{\Vert Q_y \Vert}$.  The maximizing $V$ is unchanged if we replace the problem by
\begin{align*}
&\max_{V: \tilde Q_y V  = \tilde Q_y} \frac{1}{\Vert Q_y \Vert} \sum_{a} Q_y(a) \sum_{b}  -V(b|a) \log V(b|a) \\
&=\max_{V: \tilde Q_y V  = \tilde Q_y} H(V | \tilde Q_y ).
\end{align*}
We now use the proof of property 1 to allow us to conclude that setting the rows of $V$ to be $\tilde Q_y$ solves this maximization.  Using the definition of $Y$ to see that $\Vert Q_y \Vert =\mathbb{P}(Y=y)$ and substituting the maximizing $V$, equation \eqref{eqn1:hxcystep1} becomes
\begin{align*}
\sum_{a} Q_y(a) \sum_{b}  -\tilde Q_y(b) \log \tilde Q_y(b) &= \mathbb{P}(Y=y)H(\tilde Q_y) \\
&= \mathbb{P}(Y=y)H(X | Y=y)
\end{align*}
Summing over the subgraphs gives the result.
\end{IEEEproof}
\begin{property}
\label{prop:kappacont}
Let $G$ be a graph and $Q^{(n)}$ be a sequence of distributions (on the vertices of $G$) converging to distribution $Q^\infty$.  Then
\[
\limsup_{n \to \infty} \kappa(G,Q^{(n)}) \leq \kappa(G,Q^\infty)
\]
(I.e. $\kappa(G,\cdot)$ is upper semicontinuous in $Q$ for a fixed $G$.)
\end{property}
\begin{IEEEproof}
	Let
	\[
	V^{(n)} = \mathop{\argmax_{V:V \ll G}}_{Q^{(n)} V = Q^{(n)}} H(V|Q^{(n)}),
	\]
where $V^{(n)}$ exists because we are maximizing a continuous function over a compact set.   By choosing a subsequence and relabeling we may arrange it so that $H(V^{(n)}|Q^{(n)}) \to \limsup H(V^{(n)}|Q^{(n)})$ and $V^{(n)} \to V^\infty$, where both $V^\infty \ll G$ and $Q^\infty V^\infty =Q^\infty$ are true.  In which case
\begin{align*}
\limsup_{n \to \infty} \kappa(G,Q^{(n)}) &= \limsup_{n \to \infty} H(V^{(n)}|Q^{(n)}) \\
&= H(V^\infty|Q^\infty) \leq \kappa(G,Q^\infty).
\end{align*}
\end{IEEEproof}

\section{Bounding Witsenhausen's Rate}
\label{sect:wr}
We recall that in Witsenhausen's problem \cite{Witsenhausen:1976p290} the goal is communication of $X^n$ to the decoder who has access to $Y^n$ under the criterion $P_{XY}^n(X^n = \hat X^n)=1$.  This requirement is stricter than the vanishing error probability criterion of Slepian-Wolf and increases the rate from $H(X|Y)$ to $R(G_X)$.  Witsenhausen's scheme is as follows: the decoder sees $Y^n$, a realization of the side-information and can identify the set $Z(Y^n)$ and this set forms a subgraph in $G^n_X$.  If the vertices of $G_X^n$ are colored then the encoder can send this color to the decoder, which can then uniquely identify the source symbol in $Z(Y^n)$. And a result of  \cite{Witsenhausen:1976p290} proves that when encoding blocks of length $n$, $\gamma(G_X^n)$ the smallest size of the signaling set possible.

When considering very large blocklengths, the fact that there are only polynomially many types means we can send the type essentially for free.  A possible modification of Witsenhausen's scheme is as follows.  First, fix the blocklength $n$ and for every type $Q_X$, the encoder and decoder agree on a coloring of the graph $G^n_X(T^n_{Q_X})$ using $\gamma(G^n_X(T^n_{Q_X}))$ colors.  The encoder and decoder operate as follows.

\emph{Encoder:} The encoder first communicates $Q_\mathbf{x}$, the type of the source sequence.  Next the encoder looks at the graph $G^n_X(T^n_{Q_\mathbf{x}})$, that is the subgraph of $G^n_X$ induced by $T^n_{Q_\mathbf{x}}$ and sends the color of vertex $\mathbf{x}$ to the decoder.

\emph{Decoder:} The decoder sees side-information $\mathbf{y}$ and identifies the set $Z(\mathbf{y})$.  Knowing the type the decoder can examine the induced subgraph $G^n_X(T^n_{Q_\mathbf{x}} \cap Z(\mathbf{y}))$ and using the color from the encoder, identify the source sequence.

The following lemma shows that this scheme is asymptotically optimal.

\begin{lemma}
	\begin{equation}
	R(G) = \lim_{n \to \infty} \max_{Q_X \in \mathcal{P}^n(\mathcal{X})} \frac{\log \gamma(G_X^n(T^n_{Q_X}))}{n}
	\end{equation}
\end{lemma}
\begin{IEEEproof}
The number of bits used by our scheme is an upper bound on $R(G)$ and hence
\begin{align*}
R(G) &\leq \liminf_{n \to \infty} \left [ \frac{\log (n+1)^{\vert \mathcal{X} \vert}}{n} + \max_{Q_X \in \mathcal{P}^n(\mathcal{X})}\frac{\log \gamma(G_X^n(T^n_{Q_X}))}{n} \right ] \\
&= \liminf_{n \to \infty} \max_{Q_X \in \mathcal{P}^n(\mathcal{X})} \frac{\log \gamma(G_X^n(T^n_{Q_X}))}{n}
\end{align*}

But trivially we also have
\[
R(G)  \geq \limsup_{n \to \infty}\max_{Q_X \in \mathcal{P}^n(\mathcal{X})}  \frac{\log \gamma(G_X^n(T^n_{Q_X}))}{n}
\]
where we used the fact that the chromatic number of the subgraph is at most the chromatic number of $G_X$.    
\end{IEEEproof}

We now bound the chromatic number of the induced subgraph in two steps.  First we give a degree bound on induced subgraph.

\begin{lemma}
\label{lem:degsg}
	Let $Q_X \in \mathcal{P}^n(\mathcal{X})$.  Then
\begin{align}
	\label{eqn:deltabound}
(n+1)^{-\vert\mathcal{X}\vert\vert\mathcal{X}\vert}\exp(n\kappa_n(G_X,Q_X)) - 1	\le \Delta(G_X^n(T^n_{Q_X})) \leq (n+1)^{\vert\mathcal{X}\vert\vert\mathcal{X}\vert}\exp(n\kappa_n(G_X,Q_X))
\end{align}
where
\begin{align}
 \kappa_n(G_X, Q_X) = \mathop{\mathop{\max_{V: Q_X \times V \in \mathcal{P}^n}}_{V \ll G_X}}_{Q_X V = Q_X} H(V | Q_X).
\end{align}
Note: $\kappa_n$ maximizes over types rather than distributions, but of course we may replace $\kappa_n$ by $\kappa$ in the right-hand equality of \eqref{eqn:deltabound} to get another valid upper bound.
\end{lemma}
\begin{IEEEproof}
	 Suppose $\mathbf{x} \in T^n_{Q_X}$, and let $W(\mathbf{x})$ denote the neighbors of $\mathbf{x}$ in the induced subgraph $G^n_X(T^n_{Q_X})$.  We  partition the set $\{ (\mathbf{x}, \mathbf{x'}) : \mathbf{x'} \in W(\mathbf{x}) \}$ by joint type $Q_{XX'}$ and observe that each joint type can be written as $Q_{X} \times V$ for some $V$.  One may verify $V \ll G_X$.  One also sees that $Q_X V = Q_X$, since $(\mathbf{x}, \mathbf{x'}) \in T^n_{Q_{XX'}} \cap E(G^n(T^n_{Q_X}))$ implies $Q_X' = Q_X$ and writing $Q_{XX'} = Q_X \times V$, tells us that $Q_X V = Q_X$.

For any $\mathbf{x} \in T^n_{Q_X}$ we can count the number of strings in $W(\mathbf{x})$ by decomposing $\{ (\mathbf{x}, \mathbf{x'}) : \mathbf{x'} \in W(\mathbf{x'}) \}$ into joint types, choosing a $V$ for each joint type and using the standard cardinality bounds for type classes.  Thus
\begin{align*}
\Delta(G_X^n(T^n_{Q_X})) &\leq \mathop{\sum_{V:V \ll G}}_{Q_XV = Q_X} T^n_{V}(\mathbf{x}) \\
&\leq \mathop{\sum_{V:V \ll G}}_{Q_XV = Q_X} \exp(nH(V | Q_X)) \\
&\leq (n+1)^{\vert\mathcal{X}\vert\vert\mathcal{X}\vert} \mathop{\max_{V:V \ll G}}_{Q_XV = Q_X} \exp(nH(V | Q_X)).
\end{align*}
For the reverse inequality, we let $\Delta(\mathbf{x})$ denote the degree of vertex $\mathbf{x}$ in the induced subgraph.  Then
\begin{align*}
\Delta(\mathbf{x}) = \mathop{\mathop{\sum_{V: Q_X \times V \in \mathcal{P}^n}}_{V \ne I, V \ll G}}_{Q_XV = Q_X} T^n_{V}(\mathbf{x}).
\end{align*}
To see this, note first that if $V$ arises by selecting a $\mathbf{x}' \in W(\mathbf{x})$, then $T_V(\mathbf{x}) \subset W(\mathbf{x})$.  And second, that any $V \ne I$ with $V \ll G$ and $Q_X V = Q_X$ gives rise to a neighbor.  Then because $\Delta(G_X^n(T^n_{Q_X})) = \max_{\mathbf{x} \in T_{Q_X}} \Delta(\mathbf{x})$, we have
\begin{align*}
	\Delta(G_X^n(T^n_{Q_X})) &= \max_{\mathbf{x} \in T_{Q_X}} \mathop{\mathop{\sum_{V: Q_X \times V \in \mathcal{P}^n}}_{V \ne I, V \ll G}}_{Q_XV = Q_X} T^n_{V}(\mathbf{x}) \\
	\Delta(G_X^n(T^n_{Q_X})) &= \max_{\mathbf{x} \in T_{Q_X}} \mathop{\mathop{\sum_{V: Q_X \times V \in \mathcal{P}^n}}_{V \ll G}}_{Q_XV = Q_X}  T^n_{V}(\mathbf{x}) - 1.
\end{align*}
Using the cardinality bound for typeclasses we get
\begin{align*}
	\Delta(G_X^n(T^n_{Q_X})) &\ge \max_{\mathbf{x} \in T_{Q_X}} \mathop{\max_{V:V \ll G}}_{Q_XV = Q_X} T^n_{V}(\mathbf{x}) - 1 \\
	&\ge \max_{\mathbf{x}\in T_{Q_X}} (n+1)^{-\vert \mathcal{X} \vert\vert \mathcal{X} \vert}  \mathop{\max_{V:V \ll G}}_{Q_XV = Q_X} \exp(n(H(V|Q_X))) -1\\
	&= (n+1)^{-\vert \mathcal{X} \vert\vert \mathcal{X} \vert}  \mathop{\max_{V:V \ll G}}_{Q_XV = Q_X} \exp(n(H(V|Q_X))) -1
\end{align*}
where we implicitly assumed we still have $Q_X \times V \in \mathcal{P}^n$.
\end{IEEEproof}
Using the previous lemma we bound $R(G)$ as follows
\begin{theorem}
\label{thm:kappabound}
\[
R(G_X) \leq \max_{Q_X \in \mathcal{P}(\mathcal{X})} \kappa(G_X, Q_X).
\]
\end{theorem}
\begin{IEEEproof}
A well-known fact from graph theory tells us that $\gamma(G) \leq \Delta(G) + 1$ \cite[sec 5.2]{DiestelGT}.  This combined with the previous lemma gives 
\begin{align*}
&\max_{Q_X \in \mathcal{P}^n(\mathcal{X})} \frac{\log \gamma(G_X^n(T^n_{Q_X}))}{n} \\
\quad &\leq \max_{Q_X \in \mathcal{P}^n(\mathcal{X})} n^{-1}\log \left [ (n+1)^{\vert\mathcal{X}\vert\vert\mathcal{X}\vert}\exp(n\kappa_n(G_X,Q_X)) + 1\right ] \\
&\leq \max_{Q_X \in \mathcal{P}(\mathcal{X})} n^{-1}\log \left [ (n+1)^{\vert\mathcal{X}\vert\vert\mathcal{X}\vert}\exp(n\kappa(G_X,Q_X)) + 1\right ]
\end{align*}
where the final line used the fact that in both maximizations we maximize over a larger set.  Taking limits as $n \to \infty$ gives the result.
\end{IEEEproof}
We now discuss the tightness of the bound.

\subsection{Tightness of the bound in Theorem \ref{thm:kappabound}}

We note that the bound given by $\kappa$ on $R(G)$ need not be tight. To this see, consider the graph $G$ with $V(G)=\{0,1,\ldots,2^n\}$ and $E(G)=\{ (n, n+1) : n \ge 0 \} \cup \{ (0,n) : n \ge 2\}$.  It is clear that $\gamma(G)=3$ for all $n$, and hence $R(G) \leq \log 3$.  Yet, if we choose
\begin{align*}
V(b \vert 0) &= \begin{cases}
			0 		& \text{if } b=0 \\
			2^{-n} 	& \text{otherwise}
	\end{cases} \\
V(b | a \neq 0) &= \begin{cases}
			1 	&	\text{if } b = 0 \\
			0   & 	\text{otherwise}
		\end{cases} \\
Q &= \left [ \frac{1}{2}, \frac{1}{2^{n+1}}, \frac{1}{2^{n+1}} \ldots \frac{1}{2^{n+1}} \right ]
\end{align*}
one sees that $V \ll G$ and therefore that
\[
\kappa(G,Q) \geq H(V | Q) = \frac{1}{2} \log 2^n = \frac{n}{2}.
\]
Although the gap between $R(G)$ and the bound of Theorem \ref{thm:kappabound} may be arbitrarily large, note that that the bound of Theorem \ref{thm:kappabound} is a convex program, where as the computation of even $\gamma(G)$ is NP-complete.  Hence although we do not know whether our bound is ever better than the bound provided by $\gamma(G)$, from a computational point of view our bound has an advantage.

\section{Improved Exponents for Lossless Source Coding}
\label{sect:exp}
We consider the same setup as in Figure \ref{fig:scsi}.  The encoder/decoder pair are functions $\psi:\mathcal{X}^n \to \mathcal{M}$ and $\varphi:\mathcal{M} \times \mathcal{Y}^n \to \mathcal{\hat X}^n$, where $\mathcal{M}$ is a fixed set.   We define the error probability to be
\begin{equation}
P_e(\psi, \varphi) = \Pr(X^n \ne \hat X^n)
\end{equation}
where $\hat X^n = \varphi(\psi(X^n),Y^n)$.  In this section we are interested in the asymptotic behaviour of the error probability $P_e(\psi,\varphi,
\Delta)$ as $n$ gets large.  
We define the \emph{error exponent} (or \emph{reliablity function}) to be
\begin{equation}
\theta(R,P_{XY}) = \lim_{\epsilon \downarrow 0} \liminf_{n \to \infty} -\frac{1}{n} \log \left [ \min_{(\psi,
\varphi)} P_e(\psi,\varphi) \right ]
\end{equation}
where the minimization ranges over all encoder/decoder pairs satisfying \begin{equation}
\label{rateconst}
\frac{1}{n} \log | \mathcal{M} | \leq R + \epsilon.
\end{equation}
Our main result is
\begin{theorem}
\label{thm:exponent}
For any $R>0$ and $P_{XY} \in \mathcal{P}(\mathcal{X} \times \mathcal{Y})$,
\begin{align}
\notag\theta(R,P_{XY}) \geq
 \mathop{\inf_{Q_{XY}: }}_{\min(\kappa(G_X,Q_X),\log \gamma(G_X)) \ge R} & \Big [ D(Q_{XY} || P_{XY})  \\
&\qquad + (R - H_Q(X|Y))^+ \Big ]
\end{align}
where $G_X$ is the characteristic graph of the source $P_{XY}$.
\end{theorem}

To achieve this exponent we use the following scheme.  First, fix the blocklength $n$.  For every type $Q_X$, the encoder and decoder agree on a coloring of the graph $G^n_X(T^n_{Q_X})$ using $\gamma(G^n_X(T^n_{Q_X}))$ colors.  When $
\log \gamma(G^n_X(T^n_{Q_X})) \ge nR$, the encoder and decoder agree on a random binning of the typeclass $T^n(Q_X)$ into $\exp(nR)$ bins.  The encoder's message set is
\begin{align*}
&\mathcal{M}= \mathcal{M}_1 \times \mathcal{M}_2 \text{ where } \\
&\mathcal{M}_1 = \{1,2,\ldots,\exp(nR) \}, ~ \mathcal{M}_2 = \{1,2,\ldots,(n+1)^{|\mathcal{X}|} \}
\end{align*}

\emph{Encoder:} The encoder sends the type $Q_\mathbf{x}$ of the string. If $\log \gamma(G^n_X(T^n_{Q_\mathbf{x}})) < nR$, then there is sufficient rate to send the color to the decoder.  If not, the encoder sends the bin index of the string $\mathbf{x}$.  In both cases we let $U(\mathbf{x})$ denote the index sent to the decoder.

\emph{Decoder:} The decoder receives the index of the type and side information $\mathbf{y}$.  If $\log \gamma(G^n_X(T^n_{Q_X})) < nR$ the color index and the side information allow the decoder to reproduce $X^n$ without error.  In the opposite case, the decoder receives a bin index, looks in that bin and chooses an $\mathbf{x}$ in the bin so that $H(\mathbf{x} | \mathbf{y}) \le H(\mathbf{\tilde x} | \mathbf{y})$ for all other $\mathbf{\tilde x}$ in the bin.

\subsection{Analysis}

To prove our theorem, we will use the following definition and lemmas.  Let
$$\mathcal{E} = \{ (\mathbf{x},\mathbf{y}) : \log \gamma(G_X^n(T^n_{ Q_\mathbf{x}})) \ge nR \}. $$
Observe on $\mathcal{E}^c$ our scheme makes no error.
\begin{lemma}
\label{lem:empent}
For all strings $\mathbf{x},\mathbf{y}$, let $$S(\mathbf{x} | \mathbf{y}) = \{ \mathbf{\tilde  x} | 
H(\mathbf{\tilde x} | \mathbf y) \leq H(\mathbf x | \mathbf y), Q_{\mathbf{\tilde x}} = Q_\mathbf{x} \}.$$  Then $$|S(\mathbf{x}|\mathbf{y})| \leq (n
+1)^{|\mathcal{X}||\mathcal{Y}|} \exp(nH(\mathbf{x}|\mathbf{y})).$$
\end{lemma}

\begin{IEEEproof}
\begin{align*}
|S(\mathbf x | \mathbf y)| & \leq \vert	\{ \mathbf{\tilde  x} | 
	H(\mathbf{\tilde x} | \mathbf y) \leq H(\mathbf x | \mathbf y) \} \vert \\
&= \sum_{V: V \in \mathcal{C}^n(Q_\mathbf{y},\mathcal{X})}\sum_{\tilde 
{\mathbf{x}} \in 
T_V(\mathbf{y}): H(\tilde{\mathbf{x}}|\mathbf{y}) \le H(\mathbf{x}|\mathbf{y})}1 \\
&= \mathop{\sum_{V: V \in \mathcal{C}^n(Q_\mathbf{y},\mathcal{X})}}_{ H(V|Q_\mathbf{y}) \leq 
H(\mathbf{x}|\mathbf{y})}  |T_V(\mathbf{y})| \\
&\leq  \mathop{\sum_{V: V \in \mathcal{C}^n(Q_\mathbf{y},\mathcal{X})}}_{ H(V|Q_\mathbf{y})\leq 
H(\mathbf{x}|\mathbf{y})} \exp(nH(\mathbf{x}|\mathbf{y})) \\
&\leq (n+1)^{|\mathcal{X}||\mathcal{Y}|}
\exp(nH(\mathbf{x}|\mathbf{y}))
\end{align*}
\end{IEEEproof}

\begin{lemma}
\label{lem:binerror}
For all strings $\mathbf{x},\mathbf{y}$
\[
P(X^n \ne \hat X^n | X^n = \mathbf{x}, Y^n = \mathbf{y}) \leq \exp(-n(R - H(\mathbf{x} | \mathbf{y}) -\delta_n)^{+}). 
\]
where $\delta_n \to 0$ with $n$.  Moreover if $(\mathbf{x},\mathbf{y})  \in \mathcal{E}^c$ then
\[
P(X^n \ne \hat X^n | X^n = \mathbf{x}, Y^n = \mathbf{y}) = 0.
\]
\end{lemma}
\begin{IEEEproof}
As noted in the specification of the decoder, for types $Q_X$ so that $\log \gamma(G^n_X(T^n_{Q_X})) < nR$ the decoder makes no error.  For the opposite case we bound the set of candidate $\tilde{\mathbf{x}}$ with $S(\mathbf{x}|\mathbf{y})$ yielding
\begin{align*}
P(X^n &\ne \hat X^n | X^n = \mathbf{x}, Y^n = \mathbf{y}) \\
&\leq \sum_{\mathbf{\tilde x} \in  S(\mathbf{x}|\mathbf{y})} P(U(\mathbf{x}) = U(\mathbf{\tilde x})) \\
&\leq (n+1)^{|\mathcal{X}||\mathcal{Y}|}\exp(-n(R - H(\mathbf{x} | \mathbf{y}))) \\
&\leq \exp(-n(R - H(\mathbf{x} | \mathbf{y}) - \delta_n))
\end{align*}
Using the fact that $P(X^n \ne \hat X^n | X^n = \mathbf{x}, Y^n = \mathbf{y}) \le 1$ gives the result.
\end{IEEEproof}
\begin{lemma}
	\label{lem:cont}
	Let $G$ be a graph, $\delta_n>0, \tilde{\delta}_n > 0, \tilde{\tilde{\delta}}_n $ sequences converging to zero,  
\begin{align*}
& F_n(Q_{XY}) = \\
	\notag &\begin{cases} D(Q_{XY} || P_{XY})  & \text{if } \kappa(G,Q_X) \ge R - \tilde{\delta}_n \\
	\   + (R- H_Q(X|Y) - \delta_n)^+  - \tilde{\tilde{\delta}}_n  &  \\
	\infty & \text{otherwise,}
	\end{cases} \\
	& F(Q_{XY})  = \\ 
	&\begin{cases} D(Q_{XY} || P_{XY}) + (R- H_Q(X|Y))^+  & \text{if } \kappa(G,Q_X) \ge R \\
	\infty & \text{otherwise}
	\end{cases}
\end{align*}
and $Q_{XY}^{(n)}$ be a sequence of distributions converging to $Q_{XY}^\infty$.  Then
\begin{equation}
\label{eqn:cont}
\liminf_{n \to \infty} F_n(Q_{XY}^{(n)}) \ge F(Q^\infty_{XY})
\end{equation}
\end{lemma}
\begin{IEEEproof}
We proceed by cases.  Case 1: $Q_{XY}^\infty$ is such that $\kappa(G,Q_X^\infty) \ge R$.  If $\kappa(G,Q_X^{(n)}) < R - \tilde{\delta}_n$ for all sufficiently large $n$, then the left-hand side is infinity and the result trivially holds.  Otherwise we appeal to the semicontinuity of the information measures.  
	
	Case 2: $Q_{XY}^\infty$ is such that $\kappa(G,Q_X^\infty) <R$.  In this case we see, by appealing to $\kappa$ property \ref{prop:kappacont}, that $\limsup \kappa(G,Q_X^{(n)}) < R$, whence \eqref{eqn:cont} holds with equality eventually.
\end{IEEEproof}

\begin{IEEEproof}[Proof of Theorem \ref{thm:exponent}]
	For any $\epsilon > 0$, we note that for sufficiently large $n$ the constraint \eqref{rateconst} is met.  
	Let $$\mathcal{T}^n = \{ Q_{XY} \in \mathcal{P}^n(\mathcal{X} \times \mathcal{Y}): \log \gamma(G_X^n(T^n_{Q_X})) \ge nR \}.$$ 
	We begin by partitioning the sequence space by joint type and computing the error probability for each type
	\begin{align*}
		P_e &= \sum_{Q_{XY}} \sum_{(\mathbf{x},\mathbf{y}) \in T^n_{Q_{XY}}} P(X^n \ne \hat X^n, X^n=\mathbf{x},Y^n = \mathbf{y}) \\
		&\stackrel{*}{\leq} \mathop{\sum_{Q_{XY} \in \mathcal{T}^n}}_{} \sum_{(\mathbf{x},\mathbf{y}) \in T^n_{Q_{XY}}} \exp(-n(R - H(\mathbf{x} | \mathbf{y}) -\delta_n)^{+}) \\
		&\qquad \times \exp(-n(D(Q_{XY}||P_{XY}) + H(Q_{XY})) \\
		&\leq \mathop{\sum_{Q_{XY} \in \mathcal{T}^n}}_{} \exp(-n((R - H_Q(X|Y) -\delta_n)^{+} \\
		&\qquad + D(Q_{XY}||P_{XY}) )) \\
		&\leq (n+1)^{\vert \mathcal{X} \vert  \vert \mathcal{Y} \vert } \mathop{\max_{Q_{XY} \in \mathcal{T}^n}}_{} \exp(-n((R - H_Q(X|Y) -\delta_n)^{+}  \\
		&\quad + D(Q_{XY}||P_{XY}) )) 
	\end{align*}
	where in $^*$ we applied a standard identity for the probability of a sequence in $T^n_{Q_{XY}}$ and Lemma \ref{lem:binerror}.  For any $G$, $\Delta(G)+1 \ge \gamma(G)$, thus
\[
\mathcal{T}^n \subseteq \{ Q_{XY} \in \mathcal{P}^n(\mathcal{X} \times \mathcal{Y}) :  \log ( \Delta(G_X^n(T^n_{Q_X})) +1 ) \ge nR \}.
\]
 Let 
\begin{align*}
& g^n(G_X, Q_{XY}) \\
 &= \log(\exp(n[\kappa(G_X,Q_X) + n^{-1}\vert \mathcal{X} \vert^2 \log(n+1)])+1)
\end{align*}
and observe that $n^{-1} g^n(G_X, Q_{XY}) \to \kappa(G_X,Q_X)$ and let $\tilde{\delta}_n=n^{-1} g^n(G_X, Q_{XY}) -\kappa(G_X,Q_X)$.
Appealing to Lemma \ref{lem:degsg} with $\kappa$ in place of $\kappa_n$, we may further bound the set by $\{ Q_{XY} \in \mathcal{P}^n(\mathcal{X} \times \mathcal{Y}) :  g^n(G_X, Q_{XY}) \ge R \}$ i.e.
$$\mathcal{T}^n \subseteq \tilde{\mathcal{T}}^n = \{ Q_{XY} \in \mathcal{P}^n(\mathcal{X} \times \mathcal{Y}):  \kappa(G_X,Q_X) + \tilde{\delta} \ge R \}.$$
Adopting the definitions from Lemma \ref{lem:cont}, with $\tilde{\tilde{\delta}}_n=n^{-1}\vert \mathcal{X} \vert \vert \mathcal{Y} \vert \log(n+1)$ we see
\begin{align}
\label{eqn:pf1}
	&-n^{-1} \log P_e \ge \min_{Q_{XY} \in \mathcal{P}^n(\mathcal{X} \times \mathcal{Y})} F_n(Q_{XY}).
\end{align}
For each $n$, let $Q_{XY}^{(n)}$ achieve the minimum in \eqref{eqn:pf1}.  Taking a convergent subsequence and relabelling we may assume that $Q_{XY}^{(n)} \to Q_{XY}^\infty$.  Hence
\begin{align*}
\liminf_{n \to \infty} F_n(Q_{XY}^{(n)}) &\stackrel{*}{\ge} F(Q^\infty_{XY}) \\
&\ge \inf_{Q_{XY} \in \mathcal{P}(\mathcal{X} \times \mathcal{Y})} F(Q_{XY})
\end{align*}
where $^*$ follows from Lemma \ref{lem:cont}.   The inequality
\[
\log \gamma(G_X) \ge n^{-1} \log(\gamma(G_X^n)) \ge n^{-1} \log (G_X^n(T_{Q_X}^n))
\]
implies that we may repeat the argument above to yield the achievable exponent
\[
\begin{cases} D(Q_{XY} || P_{XY}) + (R- H_Q(X|Y))^+  & \text{if } \log \gamma(G_X) \ge R \\
\infty & \text{otherwise}
\end{cases}
\]
Taking the maximum of both exponents gives the result.
\end{IEEEproof}

\subsection{Examples}

\begin{figure}
\centering
\scalebox{0.5}{\includegraphics*{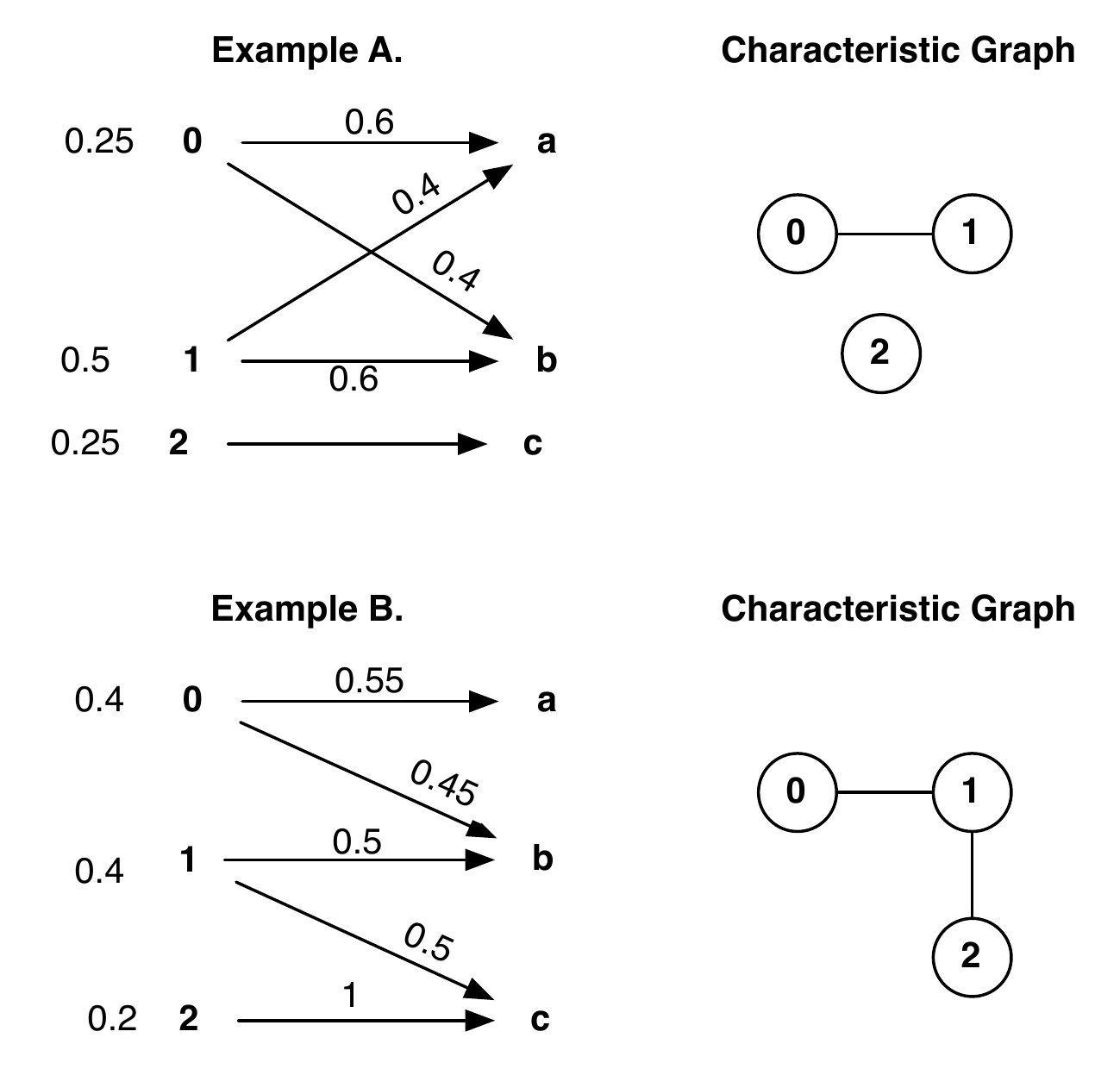}}
\caption{Two example source distributions and their characteristic graphs}
\label{fig:egs}
\end{figure}

In this section we compute the exponent of Theorem \ref{thm:exponent} and compare it with the best previously known exponents.  First we demonstrate a case in which the exponent of Theorem \ref{thm:exponent} achieves the sphere packing exponent.

When the side information is a deterministic function of the source, i.e. $Y=f(X)$, $\kappa$ property \ref{prop:hxcy} allows us to compute $\kappa$ explicitly and the optimization forces the inner most optimization to yield $Q_{Y|X}=P_{Y|X}$, i.e. the `deterministic' side information.  If we associate a $y$ to each fully connected subgraph in $G_X$, then we see that
\begin{align*}
\log \gamma(G_X) &= \max_{y \in \mathcal{Y}} \log \vert f^{-1}(y) \vert  \\
&\ge \max_{y \in \mathcal{Y}} H(X|Y=y) \\
&\ge H(X|Y).
\end{align*}
From these observations it follows that the exponent reduces to
\[
e_{SP}(R,P_{XY}) = \inf_{Q_{XY}: H_Q(X|Y) \ge R} D(Q_{XY} || P_{XY}),
\]
the sphere packing exponent for this problem.  Thus our scheme is optimal for all rates and the reliability function is determined for this problem.

\begin{figure}
\begin{center}
\scalebox{0.7}{\includegraphics*{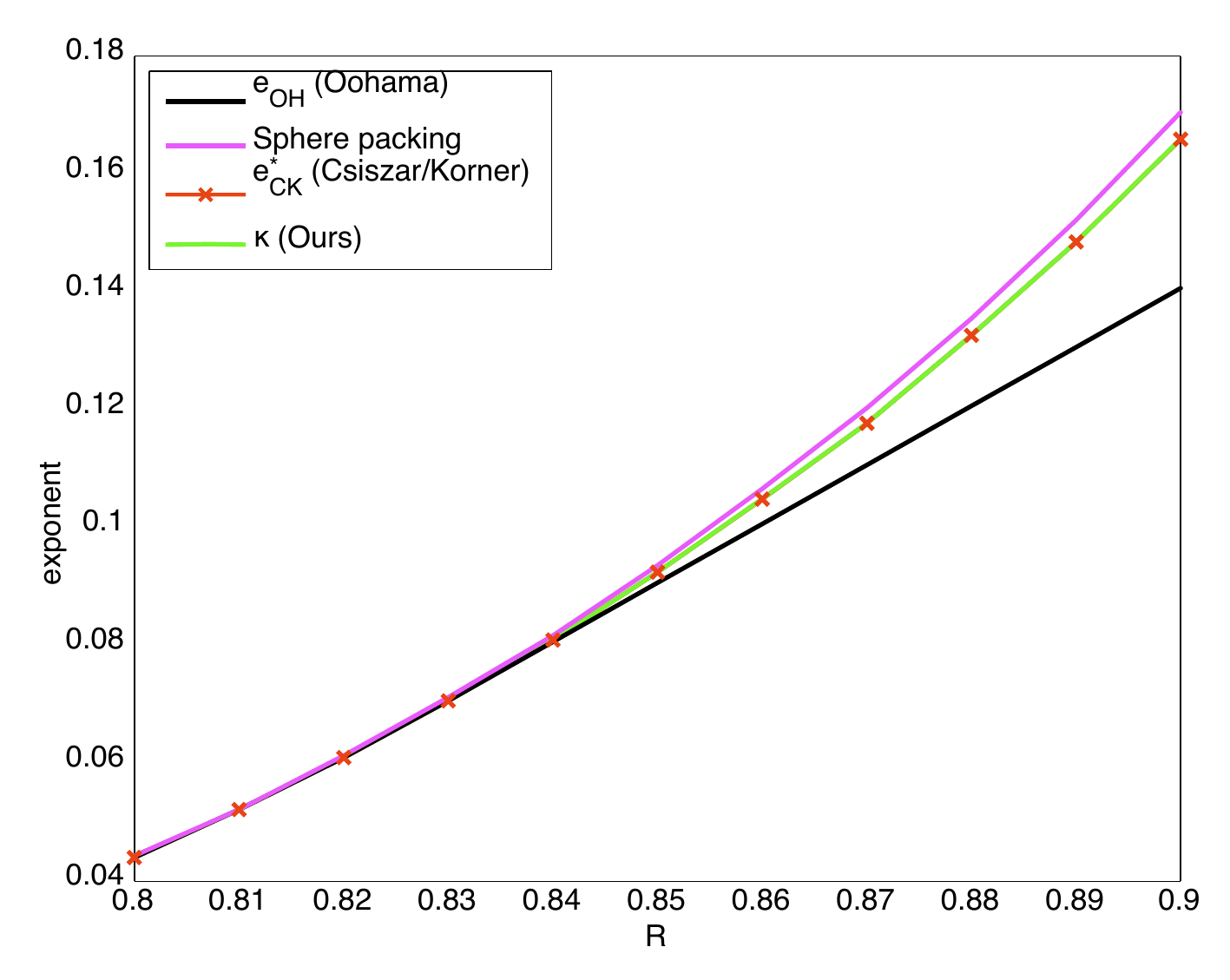}}
\end{center}
\caption{Comparing exponents for Example A of Figure \ref{fig:egs}.  Our exponent coincides with $e_{CK}^*$ and both lie below the sphere packing exponent.}
\label{fig:expplot2}
\end{figure}

For comparison with previous results we turn to \emph{Example A}\footnote{Please note the plot and discussion concerning Example A reported in a preliminary version of this work \cite{Kelly:2009p3712} were incorrect.} (see Fig. \ref{fig:egs}).  In Figure \ref{fig:expplot2} we plot our exponent against $e_{CK}^*=\max(e_{CK},e_{CK,r})$ and $e_{OH}$, where $e_{CK}$ and $e_{CK,r}$  are the expurgated and random coding exponents of Csisz\'{a}r and K\"{o}rner \cite{Csiszar:1981p220}, and $e_{OH}$ is the exponent of Oohama and Han \cite{Oohama:1994p222}.
\begin{align*}
	e_{CK} &= \inf_{Q_X} D(Q_X || P_X) \\
	&\quad + \left [ \mathop{\inf_{Q_{\tilde X X}:H(\tilde X | X) \ge R}}_{Q_{\tilde X} = Q_{X}} \rve [d_P(X, \tilde X)] + R -  H(\tilde X | X) \right ]
\end{align*}
where
\[
d_P(x, \tilde x) = - \log \left ( \sum_{y} \sqrt{P(y|x)P(y|\tilde x)}\right ).
\]
and
\begin{align*}
	e_{OH} &= \inf_{Q_{XY}: H(Q_X) \ge R} D(Q_{XY} || P_{XY}) + (R-H_Q(X|Y))^{+}.
\end{align*}

From Figure \ref{fig:expplot2} we see that our exponent lies below the sphere packing exponent and above the random coding exponent of Oohama and Han.  When compared with $e_{CK}^*$, we see that our exponent agrees (numerically) and has the benefit of semi universality.  

\begin{figure}
\begin{center}
\scalebox{0.7}{\includegraphics*{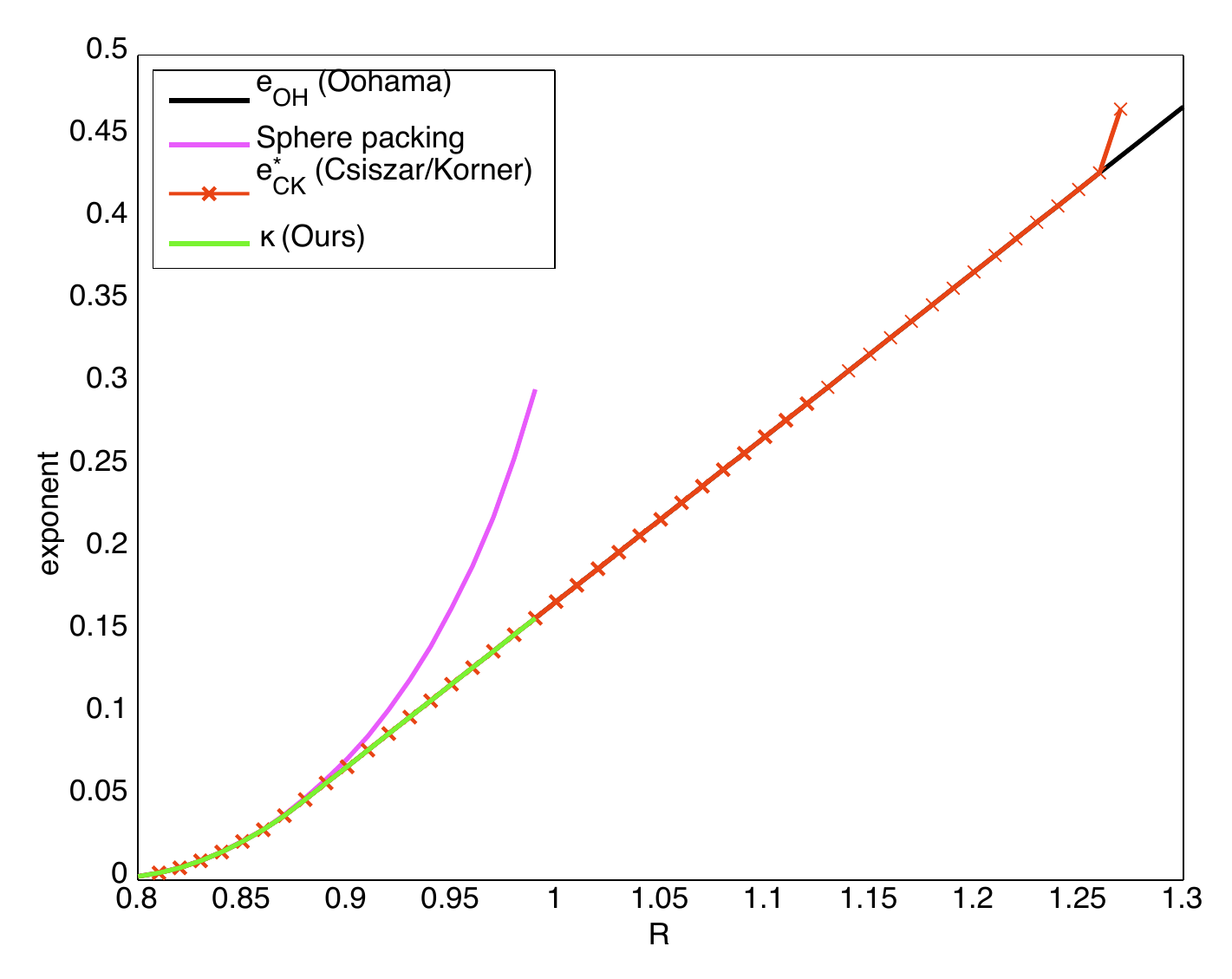}}
\end{center}
\caption{Comparing exponents for Example B of Figure \ref{fig:egs}.  Our exponent is infinite for all rates above 1 bit.  $e_{CK}^*$ is finite for some rates above 1 bit.}
\label{fig:expplotb}
\end{figure}

For \emph{Example B} (Fig. \ref{fig:egs}), it is clear that any rates in excess of one bit allows the decoder to determine the source sequence without error.  The various error exponents are plotted in Fig \ref{fig:expplotb}.  For this example our exponent is infinite for all rates above 1 bit since $\log (\gamma(G_X))=1$.  However $e_{CK}^*$ is finite for some rates above one bit, and therefore we beat $e_{CK}^*$.  Below 1 bit, $e_{OH}$, $e_{CK}^*$ and our exponent appear to agree.  The random coding exponent remains finite for all rates below $\log(3)$ bits.

\emph{Note 1:} Formally, the strongest results of \cite{Csiszar:1981p220} are obtained by using ML decoding in their equation (41) but the complexity of the optimization make computation infeasible, even for these simple examples and exploiting convexity.  However, in the particular case of our Example B, we note that if for some $R$ the exponent $e_{CK}$ is finite, then there exists a $Q_X$ for which 
\[
\mathop{\inf_{Q_{\tilde X X}:H(\tilde X | X) \ge R}}_{Q_{\tilde X} = Q_{X}} \rve [d_P(X, \tilde X)] + R -  H(\tilde X | X) < \infty.
\]
Then according to \cite[Lemma 4]{Csiszar:1981p220}, the random variables in their set $\mathcal{P}(Q_{Y|X},\tilde {Q}_{Y|X}, Q, R)$, which give equality in their equation (28) would give rise to the exponent in their equation (16) being finite.  As $e_{CK}$ is finite for some rates above 1 bit, their exponent (41) would be finite, thus at least for Example B, our exponent is strictly better than the previously known best exponent.

\emph{Note 2:} In general one also sees (via Property 1) that our exponent is never worse than the Oohama and Han exponent, because by $\kappa$ property \ref{prop:kapent} nature is forced to optimize over a smaller set of distributions.  Put another way, compared to the Oohama and Han exponent, we are able to `expurgate' more types.

\section{Improved Exponents for Wyner-Ziv}
\label{sect:coveringbinning}

When dealing with lossy reproduction it is often convenient to use `covering' (i.e. quantization) followed by binning and in this section we describe how use of the characteristic graph can yield improved error exponents in such scenarios.  We focus on lossy compression with side information i.e.  Wyner-Ziv \cite{wynerziv}.  Formally the error exponent problem in this case is as follows.

Let $\mathcal{\hat X}$ be the reproduction alphabet and $d : \mathcal{X} \to \mathcal{\hat X}$ a single letter distortion measure. Define the distortion between two strings as $d(\mathbf{x}, \hat{\mathbf{x}}) = \frac{1}{n}\sum_{i=1}^n d(x_i,\hat x_i)$.  The encoder/decoder pair are functions $f^n:\mathcal{X}^n \to \mathcal{M}$ and $g^n:\mathcal{M} \times \mathcal{Y}^n \to \mathcal{\hat X}^n$, where $\mathcal{M}$ is a fixed set.

Let $\hat X^n =g^n(f^n(X^n),Y^n)$ be the decoder's output and define the error probability 

\begin{equation} 
	P_e (f^n, g^n,\Delta, d) = \Pr\left (d (X^n, \hat X^n) > \Delta \right).
\end{equation} 
We define the Wyner-Ziv error exponent to be 

\begin{equation}
\pi(R,\Delta,P_{XY}, d) 
   = \lim_{\epsilon \downarrow 0} 
    \liminf_{n \to \infty} -\frac{1}{n} \log \left [ \min_{(f^n,
         g^n)} P_e(f^n,g^n, \Delta, d) \right ]
\end{equation}
where the minimization ranges over all encoder/decoder pairs satisfying 
\begin{equation}
\label{eqn:rateconstwz}
\log | \mathcal{M} | \leq n(R + \epsilon).
\end{equation}

Before we state the result we define another graph functional.  
\begin{definition}
\label{eqn:kappa2}
\[
\kappa_2(P_{XY}, Q_{XYU}) = [\kappa(G_{U},Q_{U}) - H(Q_{U|X}|Q_X)]^+,
\]
where the graph $G_U$ is defined from the distribution
\[
Q_{UY}(u,y) = \sum_{x \in \mathcal{X}} P_{XY}(x,y)Q_{U|X}(u|x).
\]
Note: Since $P_{XY}$ will be fixed throughout, we will abbreviate to $\kappa_2(Q_{XYU})$ or even simply $\kappa_2(Q_X)$.
\end{definition}
Our first result in this section is Theorem \ref{thm:expwz}.

	\begin{theorem}
	\label{thm:expwz}
	Let $P_{XY} \in \mathcal{P}(\mathcal{X} \times \mathcal{Y})$ and $R>0$, $\Delta > 0$, $d(\cdot,\cdot)$ be given.  Then
	\begin{align*}
\pi(R&,\Delta,P_{XY}, d) \geq \inf_{Q_X} \sup_{Q_{U|X}} \inf_{Q_Y} \sup_{\phi \in \mathcal{F}} \inf_{Q_{XYU}} \eta(R,P_{XY},Q_{XYU},\phi) 
\end{align*}
where
\begin{align*}
		\eta(R,P_{XY},Q_{XYU},\phi) &= \begin{cases}
		D(Q_{XYU}||P_{XY}Q_{U|X})  & \text{ if }\mathbb{E}_Q[d(X,\phi(Y,U))] \ge \Delta \\
		D(Q_{XYU}||P_{XY}Q_{U|X})   & \text{ if } \mathbb{E}_Q[d(X,\phi(Y,U))] < \Delta \\
		\quad + [R-I_Q(X;U)+I_Q(Y;U) ]^+ &  \quad \text{ and }\kappa_2(P_{XY},Q_{XYU}) \ge R  \\
	 	\infty & \text{ otherwise}
	\end{cases} \\
\end{align*}
and $\mathcal{F} = \{ \phi | \phi: \mathcal{Y} \times \mathcal{U} \to \hat
{\mathcal{X}} \}$. Note in the final minimization over $Q_{XYU}$, $Q_{XU}$ and $Q_Y$ are fixed to be those specified earlier in the optimization.
\end{theorem}

\emph{Discussion of Result}

In \cite{kellywagnerscexponents}, the present authors determined an achievable exponent for the Wyner-Ziv problem, obtained by replacing $\eta$ in Theorem \ref{thm:expwz} with
\[
\eta_D(R,P_{XY},Q_{XYU},\phi)  =
\begin{cases}
D(Q_{XYU}||P_{XY}Q_{U|X})  & \text{ if }\mathbb{E}_Q[d(X,\phi(Y,U))] \ge \Delta \\
D(Q_{XYU}||P_{XY}Q_{U|X})    & \text{ if } \mathbb{E}_Q[d(X,\phi(Y,U))] < \Delta \\
\quad + [R-I_Q(X;U)+I_Q(Y;U) ]^+ &  \quad \text{ and } I(X;U) \ge R  \\
\infty & \text{ otherwise,}
\end{cases}
\]
the difference being the conditions under which we switch from case 2 to case 3.  Theorem \ref{thm:expwz} is obtained by modifying the scheme in \cite{kellywagnerscexponents} taking into account the graph-based expurgation established in the previous section.  Recalling $\kappa$ property \ref{prop:kapent} we have the following inequality
\begin{align*}
	\kappa_2(Q_{XYU}) &= [\kappa(G_U,Q_{U}) - H(U|X)]^+ \\
	&\le [H(U) - H(U|X)]^+ \\
	&=I(X;U)
\end{align*}
therefore for any $R,P_{XY},\phi$ and $Q_{XYU}$ we see that $\eta_D(R,P_{XY},Q_{XYU},\phi) \le \eta(R,P_{XY},Q_{XYU},\phi)$ and the present modification yields an achievable exponent that is never any worse than the result of \cite{kellywagnerscexponents}.

\subsection{Sketch of Scheme}

Operating at blocks of length $n$, for each type $Q_X$, a test channel $Q_{U|X}^*(Q_X)=Q_{U^*|X}$ is selected.  The test channel is used to generate a codebook, $B^n(Q_X)$, of approximately $2^{nI(U^*;X)}$ codewords.  The key insight is that the (random) graph $B^n(Q_X) \cap G_{U^*}^n$, constructed from 
\[
Q_{U^*Y}(u,y) = \sum_{x \in \mathcal{X}} P_{XY}(x,y)Q_{U^*|X}(u|x)
\]
plays the same role in this problem as did the graph characteristic graph of the source $P_{XY}$ in the Slepian-Wolf problem.   

In this modified scheme, the encoder first communicates the type of $X^n$ and then if there is sufficient rate, i.e. $nR > \log \gamma(B^n(Q_X) \cap G_{U^*}^n)$, rather than communicating a bin index the encoder may send the color of the codeword in the graph $G_{U^*}$.  If there is insufficient rate, then the encoder communicates a bin index of the codeword.  For each pair marginal types $(Q_X, Q_Y)$ the decoder can choose an estimation function $\phi$ and depending on the case, either decodes using the graph, or a minimum empirical entropy decoder.  The estimation function is then used to combine the side information and the codeword to yield the reproduction.  

\subsection{Deterministic Side Information}

We now use the result of Theorem \ref{thm:expwz} to determine the reliability function when the side information is a deterministic function of the source, i.e. $Y=f(X)$ a.s. for a deterministic $f$.  We first note that in this case, the solution to the inner-most optimization must be $Q_{Y|XU}=P_{Y|X}$ else the exponent is infinite.  This reduces the problem to
\[
\inf_{Q_X} \sup_{Q_{U|X},\phi} \eta(R,P_{XY},Q_{XYU},\phi)
\]
where the distribution of $Q_{XYU}$ is $Q_X P_{Y|X} Q_{X|U}$, i.e. $U,X$ and $Y$ form a Markov chain in that order.  We can massage the exponent $\inf_{Q_X} \sup_{Q_{U|X},\phi} \eta(R,P_{XY},Q_{XYU},\phi)$ as follows
\begin{align*}
	&\inf_{Q_X} \sup_{Q_{U|X},\phi} \begin{cases}
	D(Q_{XYU}||P_{XY}Q_{U|X})  & \text{ if }\mathbb{E}_Q[d(X,\phi(Y,U))] \ge \Delta \\
	D(Q_{XYU}||P_{XY}Q_{U|X})  +  & \text{ if } \mathbb{E}_Q[d(X,\phi(Y,U))] < \Delta \\
	\quad [R-I_Q(X;U)+I_Q(Y;U) ]^+ &  \text{ and }\kappa_2(Q_{XYU}) \ge R  \\
 	\infty & \text{ otherwise}
\end{cases} \\
	&\ge \inf_{Q_X} \sup_{Q_{U|X}:Y=\nu(U),\phi} \begin{cases}
	D(Q_{XYU}||P_{XY}Q_{U|X})  & \text{ if }\mathbb{E}_Q[d(X,\phi(Y,U))] \ge \Delta \\
	D(Q_{XYU}||P_{XY}Q_{U|X})  +  & \text{ if } \mathbb{E}_Q[d(X,\phi(Y,U))] < \Delta \\
	\quad [R-I_Q(X;U)+I_Q(Y;U) ]^+ &  \text{ and } [H(U|Y) - H(U|X)]^+ \ge R  \\
 	\infty & \text{ otherwise}
\end{cases}	\\
\end{align*}
where the previous inequality follows because we maximize over a smaller set. The notation $Q_{U|X}:Y=\nu(U)$ means we consider only those test channels that result in $Y$ being a deterministic function $\nu$ of $U$. By construction $U, X$ and $Y$ still form a Markov chain in that order, thus $H(U|X) = H(U|XY)$ and we can continue the chain of equalities with
\begin{align*}
		&= \inf_{Q_X} \sup_{Q_{U|X}:Y=\nu(U),\phi} \begin{cases}
		D(Q_{XYU}||P_{XY}Q_{U|X})  & \text{ if }\mathbb{E}_Q[d(X,\phi(Y,U))] \ge \Delta \\
		D(Q_{XYU}||P_{XY}Q_{U|X})  +  & \text{ if } \mathbb{E}_Q[d(X,\phi(Y,U))] < \Delta \\
		\quad [R-I_Q(X;U|Y) ]^+ &  \text{ and } I(X;U|Y) \ge R  \\
	 	\infty & \text{ otherwise.}
	\end{cases}
\end{align*}
Note now that the only difference between $Q_{XYU}$ and $P_{XY}Q_{U|X}$ occurs in $Q_X$, so it follows that the quantity above can be written as
\begin{align*}
	&=  \inf_{Q_X} \sup_{Q_{U|X}:Y=\nu(U),\phi} \begin{cases}
	D(Q_{X}||P_X)  & \text{ if }\mathbb{E}_Q[d(X,\phi(Y,U))] \ge \Delta \text{ or } I(X;U|Y) \ge R\\
 	\infty & \text{ otherwise.}
\end{cases} \\
&=  \inf_{Q_X} \sup_{Q_{U|X},\phi} \begin{cases}
D(Q_{X}||P_X)  & \text{ if }\mathbb{E}_Q[d(X,\phi(Y,U))] \ge \Delta \text{ or } I(X;U|Y) \ge R\\
\infty & \text{ otherwise}
\end{cases}
\end{align*}
To argue the final equality,  let $Q_X$ and $R$ be fixed.  The direction $\le$ is clear since we maximize over a larger set. For $\ge$, it suffices to show that if the optimization on the left side yields $D(Q_X||P_X)$ then so does the optimization on the right.  On account of the fact that the objective is piecewise constant (over $Q_{U|X}$ and $\phi$), when the left side is finite, there exists a $Q^*_{U|X}:Y=\nu(U)$ and $\phi$ causing evaluation to $D(Q_X||P_X)$.  Suppose by way of contradiction there exists a non-deterministic $Q_{U|X}$ which yields an infinite exponent.  This means that
\[
I(X;U|Y) < R \text{ and } \mathbb{E}_Q[d(X,\phi(Y,U))] < \Delta
\]
but then by Lemma $\ref{lem:detqux}$ (which follows) we can find a deterministic $Q_{\tilde U|X}$ and corresponding $\tilde \phi$ with the property that
\[
I(X;\tilde U|Y) < R \text{ and } \mathbb{E}_Q[d(X,\tilde \phi(Y,\tilde U))] < \Delta
\]
implying that $Q_{\tilde U|X}$ would yield an infinite exponent, contradicting the optimality of $Q_{U|X}^*$. 

\begin{lemma}
\label{lem:detqux}
Let $Q_X$ be given and let $Y=f(X)$ with $P_{Y|X}$ denoting the induced conditional distribution.  Then for any $Q_{U|X},\phi$, there exists a $Q_{\tilde U|X}$ and $\tilde \phi$ so that when $Q_{XYU}=Q_X Q_{U|X}P_{Y|X}$,
\begin{align*}
1) ~ \mathbb{E}_{Q_{XYU}}[d(X,\phi(Y,U))] &= \mathbb{E}_{Q_{XY\tilde U}}[d(X,\tilde \phi(Y,\tilde U))], \\
2) ~ I(X;U|Y) &= I(X;\tilde U | Y)
\end{align*}
and 3) $Y=\nu(\tilde U)$ for some deterministic function $\nu$.
\end{lemma}
\begin{IEEEproof}
	Define $\tilde U = (U,Y)$ and $\tilde \phi(Y,\tilde U)=\phi(Y,U)$.  Then clearly conditions 1 and 3 hold.  To see condition 2 note by the chain rule
	\[
	I(X;\tilde U |Y) = I(X;U,Y|Y) = I(X;U|Y) + I(X;Y|Y,U) = I(X;U|Y).
	\]
Finally we point out that since $Y=f(X)$ we also have $\tilde U \leftrightarrow X \leftrightarrow Y$.
\end{IEEEproof}

 Rewriting this final optimization problem as
\begin{align*}
&\inf_{Q_X} \sup_{Q_{U|X},\phi} \begin{cases}
D(Q_{X}||P_X)  & \text{ if }\mathbb{E}_Q[d(X,\phi(Y,U))] \ge \Delta \text{ or } I(X;U|Y) \ge R\\
\infty & \text{ otherwise}
\end{cases} \\
&= \inf_{Q_X: R_{WZ}(\Delta,Q_X) \ge R} D(Q_X||P_X) \\
&\le \pi(R,\Delta,P_{XY},d)
\end{align*}
where $R_{WZ}(\Delta,Q_X)$ denotes the Wyner-Ziv rate distortion function for the source with $X \sim Q_X$ and $Y=f(X)$ with distortion measure $d$.  But according to the change-of-measure argument of \cite[Theorem 4]{kellywagnerscexponents}, 
\[
\pi(R,\Delta,P_{XY},d) \le \inf_{Q_X: R_{WZ}(\Delta, Q_X) \ge R} D(Q_X||P_X).
\]
Thus our scheme is optimal in the sense that it meets the change-of-measure upper bound.
\section{Connection to Channel Coding}
\label{sect:channelcoding}
In this section we demonstrate that $\kappa$ has applications in zero-error channel coding problems. Let $G=G(W)$ be the characteristic graph of the channel $W$, and $c(G)$ denote the zero error capacity (see \cite[Section III]{Korner:1998p65} for definitions).  The independence number of a graph, denoted $\alpha(G)$, is the maximum cardinality of a set of vertices of $G$ of which no two are adjacent.   We recall that $c(G) \ge \log \alpha(G)$.  According to \cite[pg 187 (prob. 18)]{ckit}
\[
\log \alpha(G) =\max_P \mathop{\min_{P_X=P_{\tilde X}=P}}_{\rve [d_W(X,\tilde X)] < \infty} I(X ; \tilde X).
\] 
Expanding the mutual information gives 
\begin{align*}
\log \alpha(G) &=\max_P \mathop{\min_{P_X=P_{\tilde X}=P}}_{\rve [d_W(X,\tilde X)] < \infty} H(\tilde X) - H(\tilde X |X) \\
&= \max_P H(P) - \mathop{\max_{P_X=P_{\tilde X}=P}}_{\rve [d_W(X,\tilde X)] < \infty}  H(\tilde X |X). 
\end{align*}
If $P \times V = P_{X\tilde X}$ then $\rve [d_W(X,\tilde X)]  < \infty$ is equivalent to $V \ll G$.  To see this note that $\rve [d_W(X,\tilde X)] < \infty$ if for all $x, \tilde x$ s.t. $P(x,\tilde x) > 0$, there is some $y$ for which $W(y|x)W(y|\tilde x) > 0$ i.e. $(x,\tilde x) \in E(G)$.  Conversely, if $V \ll G$, then $V(x | \tilde x) > 0$ only when there is some $y$ for which $W(y|x)W(y|\tilde x) > 0$.  Hence, 
\[
\log \alpha(G) =\max_P H(P) - \kappa(G,P).
\]
Hence $\kappa$ provides a lower bound on the zero error capacity of a channel $W$.

\appendices
\section{Proof of Theorem \ref{thm:expwz}}

The key to the proof is Lemma \ref{lem:expdecayoftypes}, a bound on degree of the codebook graph which holds with exponentially high probability.  With this fact established we give a scheme for coding when the bound holds and declare an error when the bound does not.

\subsection{Codebook Construction}

	Operating on blocks of length $n$, for each type $Q_X$ choose a test channel $Q_{U^*|X}=Q_{U|X}^*(Q_X)$ and let $Q_{U^*}=Q_U^*(Q_X)$ denote the resulting induced marginal type\footnote{For brevity we will use the following conventions: The random variable $U^*$ (resp. channel $Q_{U^*|X}$) refers to the random variable (resp. channel) defined by the choice of test channel for the particular $Q_X$ under consideration.}.  The test channel is used to build a codebook $B^n(Q_X)$  as follows.  For each $\mathbf{u} \in T_{Q_{U^*}}$, flip a coin with probability of heads $$p \triangleq \exp \Big (-n \Big[H(Q_{U^*|X}|Q_X)-3\frac{\vert \mathcal{U} \vert \vert \mathcal{X} \vert \log(n+1)}{n} \Big ] \Big ),$$ and add $\mathbf{u}$ to the codebook only if the coin comes up heads.   Define the distribution
	\[
	Q_{UY}(u,y) = \sum_{x \in \mathcal{X}} P_{XY}(x,y)Q_{U^*|X}(u|x)
	\]
	and let $G_{U^*}$ be the resulting characteristic graph.  The codeword for $\mathbf{x} \in T_{Q_X}$ is chosen as follows.  If $\mathcal{G}(\mathbf{x}) \triangleq B^n(Q_X) \cap T_{Q_{U|X}^*}(\mathbf{x})$ is non-empty, choose uniformly from $\mathcal{G}(\mathbf{x})$.  If $\mathcal{G}(\mathbf{x})$ is null, choose uniformly from $B^n(Q_X)$.  We let $U(\mathbf{x})$ denote the chosen codeword.  For each codebook, we define $b_{Q_X}: B^n(Q_X) \to [1,\ldots,\exp(nR)]$ (a binning function) as follows, for all $\mathbf{u} \in B^n(Q_X)$
	\[
	\Pr(b_{Q_X}(\mathbf{u}) = i) = \exp(-nR), \text{for all } i \in [1,\ldots,\exp(nR)].
	\]
	
\subsection{Scheme}
	In Lemmas \ref{lem:badtypes} and \ref{lem:expdecayoftypes} we establish that
	\begin{align*}
	\gamma(G_{U^*} \cap B^n(Q_X) ) &\le \Delta(G_{U^*} \cap B^n(Q_X)) + 1 \\
	& \stackrel{\text{w.h.p.}}{\le} \exp(n[\kappa_2(Q_X) + \lambda_n + \tilde{\delta}_n]) + 1,
	\end{align*}
	for some $\lambda_n > 0$, $\tilde{\delta}_n \to 0$ as $n \to \infty$ and where $w.h.p$ stands for probability tending to 1 as $n \to \infty$.
	 For types $Q_{X}$ in which the above bound fails to hold, we send an error message to the decoder.  For types in which the bound holds, the scheme is as follows.  To communicate the codeword to the decoder, the encoder may either give an index into the codeword set $B^n$ or using the ideas from the improved lossless binning scheme, it can color the graph $G_{U^*}^n \cap B^n(Q_X)$ using a minimal coloring and send the color of the codeword.

	\emph{Encoder:}

	The encoder first sends $k(Q_\mathbf{x})$, the type of the source sequence $Q_\mathbf{x}$.  If $\exp(n[\kappa_2(Q_\mathbf{x})+\lambda_n + \tilde{\delta}_n]) +1 < \exp(nR)$, the encoder transmits the color of the codeword in the graph $G_{U^*} \cap B^n(Q_\mathbf{x})$.  Otherwise it sends the bin index $b_{Q_\mathbf{x}}(U(\mathbf{x}))$.  Formally, we denote the encoder by $f^n: \mathcal{X}^n \to \mathcal{M}$, where
	\[
	\mathcal{M} = [1,\ldots, (n+1)^{\vert \mathcal{X}\vert}] \times [1,\ldots,\exp(nR)] 
	\]

	\emph{Decoder:}

	The decoder receives a type index, a message and the side information $\mathbf{y}$.  If $\exp(n[\kappa_2(Q_\mathbf{x})+\lambda]) +1 < \exp(nR)$ then the codeword can be decoded without error.  In the opposite case, the decoder searches the bin for a unique codeword $\hat {\mathbf{u}}$, so that among all $\tilde {\mathbf{u}}$ in the received bin, $H(\hat{\mathbf{u}}|\mathbf{y}) < H(\tilde{ \mathbf{u}}|\mathbf{y})$.  If there is no such unique codeword, the decoder chooses $\hat{\mathbf{u}}$ uniformly at randomly from the received bin.  For each pair of types $Q_X,Q_Y$, the decoder picks an reproduction function $\phi$, and declares the output as
	\[
	\hat{\mathbf{x}} \text{ where } \mathbf{\hat{x}}_j = \phi(\hat{\mathbf{u}}_j,\mathbf{y}_j).
	\]
Thus the decoder $g^n: \mathcal{Y}^n \times \mathcal{M} \to \mathcal{\hat X}$ is specified.

	\begin{lemma}
	\label{lem:badtypes}
	Let 
	\begin{align*}
		\delta_n &=  3\frac{\vert \mathcal{U} \vert \vert \mathcal{X} \vert \log(n+1)}{n} \text{ and }
		\tilde{\delta}_n=\frac{\vert \mathcal{U}\vert\vert \mathcal{U}\vert}{n}\log(n+1)  \\
	\kappa_2^{n}(Q_X) &= \kappa_2(Q_X) + \tilde{\delta}_n \text{ and } \\
	\lambda_n &= \frac{2}{n} \log(n+1) + \delta_n.
	\end{align*}
		 Then for all $n$ sufficiently large and for all types $Q_X$,
	\begin{align*}
	&\Pr(\Delta(G_{U^*}^n \cap B^n(Q_X)) > \exp(n[\kappa_2^{n}(Q_X)   +\lambda_n ]) \\
	&\le \exp_e(-(n+1)^2).
	\end{align*}
	Note the randomness in $\Delta(G_U^n \cap B^n(Q_X))$ comes from the fact that $B^n(Q_X)$ is a random set. 
	\end{lemma}
	\begin{IEEEproof}
	Let $K= 2^{n[\kappa_2^{n}(Q_X)+\lambda_n]}$, then
	\begin{align*}
		&\Pr(\Delta(G_{U^*}^n \cap B^n(Q_X)) > K) \\
		&= \Pr (\exists \mathbf{u} \in T_{Q_{U^*}} : \mathbf{u} \in B^n(Q_X), \Delta(\mathbf{u}) > K) \\ 
		 &\le \sum_{\mathbf{u} \in T_{Q_U^*}} \Pr(\mathbf{u} \in B^n(Q_X))\Pr( \Delta(\mathbf{u}) \ge K |\mathbf{u} \in B^n(Q_X) ) \\
		&\le \sum_{\mathbf{u} \in T_{Q_U^*}}\Pr( \Delta(\mathbf{u}) \ge K |\mathbf{u} \in B^n(Q_X) ).
	\end{align*}
	Let $N(\mathbf{u})$ denote the neighbors of $\mathbf{u}$ in the graph $G_U^n$, then quantity in the previous line is upper bounded by
	\begin{align*}
		& \sum_{\mathbf{u} \in T_{Q_U^*}} \Pr \Big ( \sum_{\mathbf{v} \in N(\mathbf{u})} \mathbf{1}_{\{\mathbf{v} \in B^n\}}\ge K\Big ). %
	\end{align*}
	From the construction of the codebook, we know that for each string $\mathbf{v}$, $\mathbf{1}_{\{\mathbf{v} \in B^n\}}$ is Bernoulli with parameter $p$.  Furthermore, by Lemma \ref{lem:degsg}, we know that $\vert N(\mathbf{u}) \vert \le \exp(n[\kappa(G_U,Q_U^*) + \tilde{\delta}_n]) \triangleq J(Q_X)$.  Therefore, by bounding the number of terms in the summation, letting $D_i$ be a sequence of i.i.d. Bernoulli($p$) random variables, we have
	\begin{align*}
		&\Pr(\Delta(G_{U^*}^n \cap B^n(Q_X)) > K)  \\
		&\le \vert T_{Q_{U^*}} \vert\Pr \Big (\sum_{i=1}^{J(Q_X)} D_i \ge K\Big ).
	\end{align*}
	Focusing on the probability, using the exponential form of Markov's inequality, one has for any $\theta > 0$
	\begin{align}
	\notag	\Pr \Big (\sum_{i=1}^{J(Q_X)} D_i \ge K\Big ) &\le \frac{\exp_e( J(Q_X) \ln ( 1+ p(e^\theta -1)))}{\exp_e(\theta K)} \\
	\notag	&\le \frac{\exp_e( J(Q_X) p(e^\theta -1))}{\exp_e(\theta K)} \\
	\notag	&\le \frac{\exp_e( J(Q_X) pe^\theta )}{\exp_e(\theta K)} \\
	\label{eqn:badtypesproof} &\le \exp_e(2^{n[\kappa_2(Q_{X})+\delta_n+\tilde{\delta}_n]+\theta \log e} -\theta  2^{n[\kappa_2(Q_{X})+\tilde{\delta}_n+\lambda_n]}).
	\end{align}
	Choosing $\theta=1$, we have
	\begin{align*}
		\Pr \Big (\sum_{i=1}^{J(Q_X)} D_i \ge K\Big )	&\le \exp_e (2^{n[\kappa_2(Q_{X})+\delta_n +\tilde{\delta}_n]}(2^{\log e} - (n+1)^2)).
	\end{align*}
	For $n\ge 1$, $(e -(n+1)^2) < -1$, hence
	\begin{align*}
	\Pr(\Delta(G_{U^*}^n \cap B^n(Q_X)) > K) &\le \vert T_{Q_{U^*}} \vert \exp_e(-2^{n[\kappa_2(Q_X) + \delta_n + \tilde{\delta}_n]}) \\
	&\le \vert T_{Q_{U^*}} \vert\exp_e(-2^{n\delta_n}) \\
	&\le \vert T_{Q_{U^*}} \vert \exp_e(-(n+1)^3),
	\end{align*}
	for all $n$ sufficiently large.  Since $\vert T_{Q_{U^*}} \vert$ is only exponential in $n$, the result holds.
	\end{IEEEproof}

	On account of the previous lemma, we have a bound, which holds with high probability, on the degree of $G_{U^*} \cap B^n(Q_X)$.   For each $Q_{XYU}$, we define the event $F(Q_{XYU})$ as follows
	\[
	F(Q_{XYU}) \triangleq \{ \Delta(B^n(Q_X) \cap Q_{U^*}) > e^{n[\kappa_2^{n}(Q_X)+\lambda_n]} \}.
	\]
	\begin{lemma}
	\label{lem:expdecayoftypes}
		For all $n$ sufficiently large and any type $Q_{XYU}$
	\[
	\Pr(F(Q_{XYU})) \le \exp(-(n+1)^2).
	\]
	\end{lemma}
	\begin{IEEEproof}
	The result follows directly from Lemma \ref{lem:badtypes}.
	\end{IEEEproof}
	
	In the remainder of this appendix $\kappa_2^n$ and $\lambda_n$ will be defined as in the statement of Lemma \ref{lem:badtypes}.

\subsection{Error Analysis}
	Let
	\begin{align*}
	\mathcal{E}_1 &= \{ (\mathbf{x},\mathbf{y},\mathbf{u}): \mathbf{u} \not \in T_{Q_{U|X}^*}(\mathbf{x}) \} \\
	\mathcal{E}_2 &= \{ (\mathbf{x},\mathbf{y},\mathbf{u}): \mathbf{u} \in T_{Q_{U|X}^*}(\mathbf{x}), d(\mathbf{x},\phi_{Q_{\mathbf{x}},Q_{\mathbf{y}}}(\mathbf{u},\mathbf{y})) < \Delta \\
	&\quad \exp(n[\kappa_2^n(Q_\mathbf{x}) + \lambda_n]) + 1 \ge \exp(nR)  \} \\
	\mathcal{E}_3 &= \{ (\mathbf{x},\mathbf{y},\mathbf{u}): \mathbf{u} \in T_{Q_{U|X}^*}(\mathbf{x}), d(\mathbf{x},\phi_{Q_{\mathbf{x}},Q_{\mathbf{y}}}(\mathbf{u},\mathbf{y})) < \Delta \\
	&\quad \exp(n[\kappa^n_2(Q_\mathbf{x}) + \lambda_n]) + 1 < \exp(nR)  \} \\
	\mathcal{E}_4 &= \{ (\mathbf{x},\mathbf{y},\mathbf{u}): \mathbf{u} \in T_{Q_{U|X}^*}(\mathbf{x}), d(\mathbf{x},\phi_{Q_{\mathbf{x}},Q_{\mathbf{y}}}(\mathbf{u},\mathbf{y})) \ge \Delta \}
	\end{align*}
	and
	\begin{align*}
	\mathcal{D}_1 &= \{ Q_{XYU}: Q_{U|X} \neq Q_{U|X}^*(Q_X) ) \} \\
	\mathcal{D}_2 &= \{ Q_{XYU}:\exp(n[\kappa^n_2(Q_X) + \lambda_n]) + 1 \ge \exp(nR) \\
	&\quad Q_{U|X} =Q_{U|X}^*(Q_X), \mathbb{E}_Q[d(X,\phi_{Q_X,Q_Y}(U,Y)) < \Delta  \} \\
	\mathcal{D}_3 &= \{ Q_{XYU}:\exp(n[\kappa^n_2(Q_X) + \lambda_n]) + 1 < \exp(nR) \\
	&\quad Q_{U|X} =Q_{U|X}^*(Q_X), \mathbb{E}_Q[d(X,\phi_{Q_X,Q_Y}(U,Y)) < \Delta  \} \\
	\mathcal{D}_4 &= \{ Q_{XYU}: Q_{U|X} =Q_{U|X}^*(Q_X), \mathbb{E}_Q[d(X,\phi_{Q_X,Q_Y}(U,Y)) \ge \Delta  \}.
	\end{align*}

The sets defined above and the following Lemmas allow us to bound the error probability for our improved scheme.

	\begin{lemma}
		\label{lem:coveringerror}
		Let $X^n, Y^n, U^n$ be generated according to our scheme, then for all $n$ sufficiently large and all $(\mathbf{x},\mathbf{y},\mathbf{u}) \in \mathcal{E}_1$
		\[
		\Pr(X^n=\mathbf{x},Y^n=\mathbf{y}, U^n=\mathbf{u}, F^c(Q_{\mathbf{xyu}})) \le \exp(-(n+1)^2).
		\]
	\end{lemma}
	\begin{IEEEproof}
		\begin{align*}
		&\Pr(X^n=\mathbf{x},Y^n=\mathbf{y}, U^n=\mathbf{u}, F^c(Q_{\mathbf{xyu}})) \\ &= 	\Pr(X^n=\mathbf{x},Y^n=\mathbf{y}, U^n=\mathbf{u}) \\ 
		&\quad \times \Pr(F^c(Q_{\mathbf{xyu}})| X^n=\mathbf{x},Y^n=\mathbf{y}, U^n=\mathbf{u}) \\
		&\le \Pr(X^n=\mathbf{x},Y^n=\mathbf{y}, U^n=\mathbf{u})
		\end{align*}
		Let $A$ denote the event that there does not exist a $\mathbf{u} \in B^n(Q_\mathbf{x})$ such that $\mathbf{u} \in T_{Q_{U^*|X}}(\mathbf{x})$. 	For $(\mathbf{x},\mathbf{y},\mathbf{u}) \in \mathcal{E}_1$, the event $\{X^n=\mathbf{x},Y^n=\mathbf{y}, U^n=\mathbf{u}\}$ implies that the event $A$ has occurred.  Hence
		\begin{align*}
		&\Pr(X^n=\mathbf{x},Y^n=\mathbf{y}, U^n=\mathbf{u})		 \\
		&= \Pr(X^n=\mathbf{x},Y^n=\mathbf{y}, U^n=\mathbf{u}, A) \\
		&\le \Pr(X^n=\mathbf{x}) \Pr(A | X^n=\mathbf{x}) \\
		&\le \Pr(A | X^n=\mathbf{x}).
		\end{align*}
	Recalling $p$ was the probability that each codeword is added to the codebook.  We have
	\begin{align*}
		\Pr(A | X^n=\mathbf{x}) &= \Pr(\forall \mathbf{u} \in T_{Q_{U^*|X}}: \mathbf{u} \not \in B^n(Q_\mathbf{x})) \\
		&= (1-p)^{\vert T_{Q_{U^*|X}(\mathbf{x})} \vert} \\
		&\le \exp(-p\vert T_{Q_{U^*|X}(\mathbf{x})} \vert).
	\end{align*}
	For $\mathbf{x} \in T_{Q_X}$ we have the lower bound,
	\[
	|T^n_{Q_{U^*|X}}(\mathbf{x})|  \geq (n+1)^{-|\mathcal{X}||\mathcal{U}|} \exp(nH(Q_{U^*|X}|Q_X))
	\]
	substituting this and the value of $p$ we get
	\begin{align*}
	\Pr(A | X^n=\mathbf{x}) &\le \exp \Big (-\exp \Big (n \Big [3\frac{\vert \mathcal{U} \vert \mathcal{X} \vert}{n} \log(n+1) - \frac{\vert \mathcal{U} \vert \mathcal{X} \vert}{n} \log(n+1) \Big ] \Big ) \Big) \\
	&\le \exp(-(n+1)^2).
	\end{align*}
	\end{IEEEproof}
	\begin{lemma}
		\label{lem:goodtypes}
		Let $\mathbf{x},\mathbf{y},\mathbf{u} \in \mathcal{E}_1^c$, then
	\begin{align*}
	&\Pr(X^n=\mathbf{x},Y^n=\mathbf{y},U^n=\mathbf{u},F^c(Q_{\mathbf{xyu}}))  \\
		&\le P_{XY}^n(\mathbf{x},\mathbf{y}) \exp(-n[H(Q_{U|X}^*(Q_\mathbf{x})|Q_\mathbf{x}) -\delta_n]),
	\end{align*}
	where
	\[
	\delta_n = 3\frac{\vert \mathcal{U} \vert \mathcal{X} \vert}{n} \log(n+1).
	\]
	\end{lemma}
	\begin{IEEEproof}
		Proceeding as in proof of Lemma \ref{lem:coveringerror}, we have
	\begin{align*}
		&\Pr(X^n=\mathbf{x},Y^n=\mathbf{y}, U^n=\mathbf{u}, F^c(Q_{\mathbf{xyu}}))  \\
		&\le \Pr(X^n=\mathbf{x},Y^n=\mathbf{y}, U^n=\mathbf{u}) \\
		&= \Pr(X^n=\mathbf{x},Y^n=\mathbf{u}) \Pr(U^n=\mathbf{u} | X^n=\mathbf{x},Y^n=\mathbf{y}).
	\end{align*}
	Conditional on $\{X^n = \mathbf{x}\}$, the event $\{U^n = \mathbf{u} \}$ is equivalent to $\{\mathbf{u} \in B^n(Q_\mathbf{x}) \} \cap \{\mathbf{u}$ was chosen among all $\tilde {\mathbf{u}} \in B^n(Q_\mathbf{x})$ with $\tilde{\mathbf{u}} \in T_{Q_{U|X}^*}(\mathbf{x})\}$.  Bounding the latter probability by 1, we have
	\begin{align*}
		&\Pr(X^n=\mathbf{x},Y^n=\mathbf{y}, U^n=\mathbf{u}, F^c(Q_{\mathbf{xyu}})) \\ 	&\le P_{XY}^n(\mathbf{x},\mathbf{y}) \exp(-n[H(Q_{U|X}^*|Q_\mathbf{x}) -3\frac{\vert \mathcal{U} \vert \mathcal{X} \vert}{n} \log(n+1)])
	\end{align*}
	\end{IEEEproof}

	\begin{lemma}
		\label{lem:typeprobabilty}
		For any $Q_{XYU} \in \mathcal{D}_1^c$ and any $P_{XY}$
		\begin{align*}
			&\sum_{(\mathbf{x},\mathbf{y},\mathbf{u}) \in T_{Q_{XYU}}} \Pr(X^n=\mathbf{x},Y^n=\mathbf{y},U^n=\mathbf{u},F^c(Q_{XYU})) \\
			&\le \exp(-n[D(Q_{XYU}||P_{XY}Q_{U|X}^*(Q_X))-\delta_n]),
		\end{align*}
		where $\delta_n$ is the same as in the statement of Lemma \ref{lem:goodtypes}.
	\end{lemma}
	\begin{IEEEproof}
		Using the bound of Lemma \ref{lem:goodtypes} and the following identity for $(\mathbf{x},\mathbf{y}) \in T_{Q_{XY}}$,
		\[
		P^n_{XY}(\mathbf{x},\mathbf{y}) = \exp(-n[D(Q_{XY}||P_{XY})+H(Q_{XY})] ),
		\]
		we have
	\begin{align}
	\notag		&\sum_{(\mathbf{x},\mathbf{y},\mathbf{u}) \in T_{Q_{XYU}}} \Pr(X^n=\mathbf{x},Y^n=\mathbf{y},U^n=\mathbf{u},F^c(Q_{XYU})) \\
	\notag	&\le \sum_{(\mathbf{x},\mathbf{y},\mathbf{u}) \in T_{Q_{XYU}}}\exp(-n[D(Q_{XY}||P_{XY}) + H(Q_{XY})\\
	\notag	&\quad  + H(Q_{U|X}|Q_X) - \delta_n] ) \\
	\notag	&\le \exp(-n[D(Q_{XY}||P_{XY})-H(Q_{U|XY}|Q_{XY}) \\
	\label{eqn:sumbound1} &\quad + H(Q_{U|X}|Q_X) - \delta_n] ).
	\end{align}
	Applying the identity 
	\begin{align*}
	& D(Q_{XY}||P_{XY}) -H(Q_{U|XY}|Q_{XY}) + H(Q_{U|X}|Q_X)  \\
	&= D(Q_{XYU} || P_{XY}Q_{U|X})
	\end{align*}
	in \eqref{eqn:sumbound1} gives the result.
	\end{IEEEproof}

	\begin{lemma}
		\label{lem:wzbinerror}
		For $n$ sufficiently large and $(\mathbf{x},\mathbf{y},\mathbf{u}) \in \mathcal{E}_2$
		\begin{align*}
		& \Pr( d(X^n, \hat X^n) > \Delta | X^n = \mathbf{x}, Y^n = \mathbf{y}, U^n = \mathbf{u}, F^c(Q_{\mathbf{xyu}})) \\
		&\le \frac{\exp(-n[R - I_{Q_{\mathbf{xyu}}}(X; U) - I_{Q_{\mathbf{xyu}}}(U;Y) -\delta_n]^+)}{1-\exp_e(-(n+1)^2)},
	\end{align*}
	where $\delta_n \to 0$ as $n \to \infty$.
	\end{lemma}
	\begin{IEEEproof}
	Let $L$ be the event that the decoder decodes the wrong codeword, i.e.
	\begin{align*}
	L &\triangleq \{ \exists \tilde {\mathbf{u}} \ne U(X^n): H(\tilde {\mathbf{u}}|\mathbf{y}) \le H(U(X^n)|\mathbf{y}), \tilde {\mathbf{u}} \in B^n(Q_{X^n}), \\
	&\qquad b_{Q_{X^n}}(U(X^n)) = b_{Q_{X^n}}(\tilde {\mathbf{u}} )  \}
	\end{align*}
	 and note that  $\{d(X^n,\hat X^n) > \Delta \} \cap \mathcal{E}_2 \subseteq L$.  We can bound the conditional probability of $L$ as follows
	\begin{align*}
		& \Pr(L | X^n = \mathbf{x}, Y^n = \mathbf{y}, U^n = \mathbf{u}, F^c(Q_{\mathbf{xyu}})) \\
		&= \frac{\Pr(L, F^c(Q_{\mathbf{xyu}}) | X^n = \mathbf{x}, Y^n = \mathbf{y}, U^n = \mathbf{u} )}{\Pr(F^c(Q_{\mathbf{xyu}}) | X^n = \mathbf{x}, Y^n = \mathbf{y}, U^n = \mathbf{u}  )} \\
		&\le \frac{\Pr(L | X^n = \mathbf{x}, Y^n = \mathbf{y}, U^n = \mathbf{u} )}{\Pr(\Delta(B^n(Q_\mathbf{x}) \cap Q_{U^*}) \le e^{n[\kappa^n_2(Q_\mathbf{x})+\lambda_n]} )}.
	\end{align*}
	We now bound the numerator.  Recalling the definition of $S(\mathbf{u}|\mathbf{y})$ from Lemma \ref{lem:empent} and invoking the union bound gives
	\begin{align*}
		& \Pr(L | X^n = \mathbf{x}, Y^n = \mathbf{y}, U^n = \mathbf{u} )  \\
		&\le \sum_{\tilde{\mathbf{u}} \in S(\mathbf{u}|\mathbf{y})} \Pr(\tilde{\mathbf{u}} \in B^n(Q_\mathbf{x}), b_{Q_\mathbf{x}}(\mathbf{u})=b_{Q_\mathbf{x}}(\tilde{\mathbf{u}})),
	\end{align*}
	and substituting the various bounds gives
	\[
	\exp(-n[R - I_{Q_{\mathbf{xyu}}}(X; U) + I_{Q_{\mathbf{xyu}}}(U;Y) -\delta_n]^+),
	\]
	where $\delta_n = 4\frac{\vert \mathcal{U}\vert \vert \mathcal{X} \vert}{n}\log(n+1)$.  To handle the denominator, by Lemma \ref{lem:badtypes} the complementary event goes to zero super exponentially as $n \to \infty$.
	\end{IEEEproof}

	\begin{lemma}
	\label{lem:contwz}
		Let $\delta_n,\tilde{\delta}_n,\tilde{\tilde{\delta}}_n,\tilde{\tilde{\tilde{\delta}}}_n$ be positive sequences converging to 0 as $n \to \infty$,
		\begin{align*}
			\eta^{n}(R,P_{XY},Q_{XYU}, \phi) &= \begin{cases}
			D(Q_{XYU}||P_{XY}Q_{U|X}) - \delta_n & \text{if }\mathbb{E}_Q[d(X,\phi(Y,U))] \ge \Delta \\
			D(Q_{XYU}||P_{XY}Q_{U|X}) - \delta_n + [R & \text{if }\mathbb{E}_Q[d(X,\phi(Y,U))] < \Delta \\
			\quad -I_Q(X;U)+I_Q(Y;U) -\tilde{\delta}_n ]^+ -\tilde{\tilde{\delta}}_n &  \text{ and }\kappa_2^{n}(Q_{X}) + \lambda_n \ge R - \tilde{\tilde{\tilde{\delta}}}_n  \\
			\infty & \text{otherwise,}
		\end{cases} \\
	\beta^n(R,\Delta,P_{XY},d) &= \min_{Q_X} \max_{Q_{U|X}} \min_{Q_Y} \max_{\phi} \min_{Q_{XYU}} \eta^{n}(R,P_{XY},Q_{XYU},\phi) \\
		\eta(R,P_{XY},Q_{XYU},\phi) &= \begin{cases}
		D(Q_{XYU}||P_{XY}Q_{U|X})  & \text{ if }\mathbb{E}_Q[d(X,\phi(Y,U))] \ge \Delta \\
		D(Q_{XYU}||P_{XY}Q_{U|X})  +  & \text{ if } \mathbb{E}_Q[d(X,\phi(Y,U))] < \Delta \\
		\quad \{R-I_Q(X;U)+I_Q(Y;U) \}^+ &  \text{ and }\kappa_2(Q_{X}) \ge R  \\
	 	\infty & \text{otherwise}
	\end{cases} \\
	\text{and } \beta(R,\Delta,P_{XY},d) &= \inf_{Q_X} \sup_{Q_{U|X}} \inf_{Q_Y} \sup_{\phi} \inf_{Q_{XYU}} \eta(R,P_{XY},Q_{XYU},\phi).
	\end{align*}
	Then
	\[
	\liminf_{n \to \infty} \beta^n(R,\Delta,P_{XY},d)  \ge \beta(R,\Delta,P_{XY},d) 
	\]
	(Note in $\beta^n$ the maximizations are over types/conditional types and in $\beta$ over distributions.)
	\end{lemma}
	\begin{IEEEproof}
		One sees that $\kappa_2^n(Q_X) + \lambda_n = \kappa_2(Q_X) + o(n)$ is upper semicontinuous in $Q_X$, with this established the proof then follows a similar proof for the Wyner-Ziv error exponent in \cite{kellywagnerscexponents}.  
	\end{IEEEproof}

	\begin{IEEEproof}[Proof of Theorem 2]
	Define
	\[
	\mathcal{E} = \{ d(X^n,\hat X^n) > \Delta \},
	\]
	then for our scheme we have
	\begin{align*}
	P_e &= \sum_{\mathbf{x},\mathbf{y},\mathbf{u}} \Pr(\mathcal{E} | X^n = \mathbf{x},Y^n=\mathbf{y},U^n = \mathbf{u}, F(Q_{\mathbf{xyu}})) \\
	&\quad \times \Pr(X^n = \mathbf{x},Y^n=\mathbf{y},U^n = \mathbf{u},F(Q_{\mathbf{xyu}})) \\
	&+ \sum_{\mathbf{x},\mathbf{y},\mathbf{u}} \Pr(\mathcal{E} | X^n = \mathbf{x},Y^n=\mathbf{y},U^n = \mathbf{u}, F^c(Q_{\mathbf{xyu}})) \\
	&\quad \times \Pr(X^n = \mathbf{x},Y^n=\mathbf{y},U^n = \mathbf{u},F^c(Q_{\mathbf{xyu}})).
	\end{align*}
	By definition, when $F$ occurs the encoder sends an error symbol, which we assume leads to the distortion constraint being violated.  Using this observation, and rewriting the above equation, first summing over types then over sequences gives
	\begin{align*}
	P_e &\le \sum_{Q_{XYU}} \sum_{\mathbf{x},\mathbf{y},\mathbf{u} \in T_{Q_{XYU}}} \Big [ \Pr(\mathcal{E} | X^n = \mathbf{x},Y^n=\mathbf{y},U^n = \mathbf{u}, F^c(Q_{XYU})) \\
	&\qquad \times \Pr(X^n = \mathbf{x},Y^n=\mathbf{y},U^n = \mathbf{u},F^c(Q_{XYU})) \Big ] \\
	&\quad + \sum_{Q_{XYU}} \vert T_{Q_{XYU}}\vert \Pr(F(Q_{XYU})).
	\end{align*}
	On account of the fact that $\Pr(F(Q_{XYU}))$ goes to zero super exponentially for any choice of $Q_{XYU}$ and the fact that there are only exponentially many sequences and  polynomially many types, the final summand can be safely ignored for the error exponent calculation.  We use $a \preceq b$ to mean that
	\[
	\limsup_{n \to \infty} \frac{1}{n} \log a \le \limsup_{n \to \infty} \frac{1}{n} \log b.
	\]

	Let $$P(\mathbf{x},\mathbf{y},\mathbf{u})=\Pr(X^n=\mathbf{x},Y^n=\mathbf{y},U^n=\mathbf{u},F^c(Q_\mathbf{xyu}))$$
	and
	$$
	P(\mathcal{E}|\mathbf{x},\mathbf{y},\mathbf{u})=\Pr(\mathcal{E} | X^n = \mathbf{x},Y^n=\mathbf{y},U^n = \mathbf{u}, F^c(Q_{\mathbf{xyu}})).
	$$
	We now group the summation according to the sets outlined at the start of this section.  This gives
	\begin{align*}
	P_e &\preceq \sum_{Q_X} \sum_{Q_Y} \Big [  \sum_{Q_{XYU} \in \mathcal{D}_1} \sum_{\mathbf{x},\mathbf{y},\mathbf{u} \in T_{Q_{XYU}}} P(\mathbf{x},\mathbf{y},\mathbf{u}) P(\mathcal{E}|\mathbf{x},\mathbf{y},\mathbf{u})  \\
	&\quad + \sum_{Q_{XYU} \in \mathcal{D}_2} \sum_{\mathbf{x},\mathbf{y},\mathbf{u} \in T_{Q_{XYU}}} P(\mathbf{x},\mathbf{y},\mathbf{u}) P(\mathcal{E}|\mathbf{x},\mathbf{y},\mathbf{u})  \\
	&\quad + \sum_{Q_{XYU} \in \mathcal{D}_3} \sum_{\mathbf{x},\mathbf{y},\mathbf{u} \in T_{Q_{XYU}}} P(\mathbf{x},\mathbf{y},\mathbf{u}) P(\mathcal{E}|\mathbf{x},\mathbf{y},\mathbf{u})  \\
	&\quad + \sum_{Q_{XYU} \in \mathcal{D}_4} \sum_{\mathbf{x},\mathbf{y},\mathbf{u} \in T_{Q_{XYU}}} P(\mathbf{x},\mathbf{y},\mathbf{u}) P(\mathcal{E}|\mathbf{x},\mathbf{y},\mathbf{u}) \Big]
	\end{align*}
	where in the inner summations over $Q_{XYU}$ on the sets $\mathcal{D}_i$, the types of $Q_X$ and $Q_Y$ are fixed to be those set by the outer summations.  On the set $\mathcal{D}_1$, Lemma \ref{lem:coveringerror} implies the quantity $P(\mathbf{x},\mathbf{y},\mathbf{u})$ decays super exponentially.  Since there are only polynomially many types and exponentially many sequences this term can therefore be safely ignored.  On the set $\mathcal{D}_3$, conditional on the event $F^c(Q_{\mathbf{xyu}})$, the codeword can be decoded without error, and hence there is no error.  Using the result of Lemmas \ref{lem:typeprobabilty} and \ref{lem:wzbinerror} we therefore have
	\begin{align}
	\notag P_e &\preceq \sum_{Q_X} \sum_{Q_Y} \Big [  \sum_{Q_{XYU} \in \mathcal{D}_2}  \exp(-n[D(Q_{XYU}||P_{XY}Q_{U|X})-\delta_n \\
	\notag &\qquad  + [R-I_Q(X;U)+I_Q(Y;U)-\tilde{\delta}_n]^+ -\tilde{\tilde{\delta}}_n])\\
	\notag&\quad  + \sum_{Q_{XYU} \in \mathcal{D}_4}  \exp(-n[D(Q_{XYU}||P_{XY}Q_{U|X})-\delta_n]) \Big ]
	\end{align}
	where $\tilde{\tilde{\delta}}_n=-\frac{1}{n}\log(1-\exp_e(-(n+1)^2)$.  Bounding the summands by their maximum value gives
	\begin{align}
	\notag P_e &\preceq \vert \mathcal{P}^n(\mathcal{X}) \vert \max_{Q_X}  \vert \mathcal{P}^n(\mathcal{Y}) \vert \max_{Q_Y} \vert \mathcal{P}^n(\mathcal{X} \times \mathcal{Y} \times \mathcal{U}) \vert \\
	\notag &\quad \times \Big [    \max_{Q_{XYU} \in \mathcal{D}_2}  \exp(-n[D(Q_{XYU}||P_{XY}Q_{U|X})-\delta_n \\
	\notag &\qquad  +[R-I_Q(X;U)+I_Q(Y;U) - \tilde{\delta}_n]^+ -\tilde{\tilde{\delta}}_n])\\
	\label{eqn:mainthm1ps1}&\quad  + \max_{Q_{XYU} \in \mathcal{D}_4}  \exp(-n[D(Q_{XYU}||P_{XY}Q_{U|X})-\delta_n]) \Big ]
	\end{align}
	Let $$\tilde{\tilde{\tilde{\delta}}}_n(Q_X)=\frac{1}{n}\log(\exp(n[\kappa_2^{n}(Q_X)+\lambda_n])+1)-(\kappa_2^{n}(Q_X)+\lambda_n)$$
	and let $\tilde{\tilde{\tilde{\delta}}}_n$ be the maximum over $Q_X \in \mathcal{P}^n(\mathcal{X})$ of $\tilde{\tilde{\tilde{\delta}}}_n(Q_X)$; it follows that $\tilde{\tilde{\tilde{\delta}}}_n \to 0$.  Adopting the definitions from the statement of Lemma \ref{lem:contwz} and using $a+b \le 2\max(a,b)$ to combine the two sums of \eqref{eqn:mainthm1ps1} gives
	\begin{align*}
	\notag P_e &\preceq 2\vert \mathcal{P}^n(\mathcal{X}) \vert \vert\mathcal{P}^n(\mathcal{Y}) \vert \vert \mathcal{P}^n(\mathcal{X} \times \mathcal{Y} \times \mathcal{U}) \vert \\
	&\quad \times \max_{Q_X} \max_{Q_Y} \max_{Q_{XYU}:Q_{U|X}=Q_{U|X}^*(Q_X)} \exp(-n[\eta^{n}(R,P_{XY}, Q_{XYU},\phi)])
	\end{align*}
	Finally, we can optimize over $Q_{U|X}^*$ and $\phi$, and move the optimizations in the exponent to give
	\begin{align*}
	\notag P_e &\preceq 2\vert \mathcal{P}^n(\mathcal{X})\vert \vert\mathcal{P}^n(\mathcal{Y}) \vert \vert \mathcal{P}^n(\mathcal{X} \times \mathcal{Y} \times \mathcal{U}) \vert \\
	&\quad \times \exp(-n[\min_{Q_X} \max_{Q_{U|X}} \min_{Q_Y} \max_{\phi} \min_{Q_{XYU}}\eta^{n}(R,P_{XY},Q_{XYU},\phi)]).
	\end{align*}
	Taking the log, dividing by $-n$ and then taking the $\liminf_{n \to \infty}$ of both sides, invoking Lemma \ref{lem:contwz} on the righthand side gives the result. 
	\end{IEEEproof}

	\bibliographystyle{IEEEtran}
	\bibliography{bib}

\end{document}